\documentclass{article}

\usepackage{arxiv}

\usepackage[utf8]{inputenc} 
\usepackage[T1]{fontenc}    
\usepackage{hyperref}       
\usepackage{url}            
\usepackage{booktabs}       
\usepackage{amsfonts}       
\usepackage{nicefrac}       
\usepackage{microtype}      
\usepackage{lscape}
\usepackage[authoryear, sort]{natbib}

\usepackage{setspace}

\usepackage{morefloats}
\usepackage{color}
\usepackage{multirow}
\usepackage{latexsym}
\usepackage{amsmath,amssymb,amsthm,graphicx}
\usepackage{dsfont}
\usepackage{enumitem}

\allowdisplaybreaks

\usepackage{colortbl}
\newcommand{\myrowcolour}{\rowcolor[gray]{0.925}}

\newtheoremstyle{thm}
{9pt}
{9pt}
{\itshape}
{}
{\bfseries}
{.}
{ }
{}
\theoremstyle{thm}

\newtheorem{theorem}{Theorem}[section]

\newtheorem{corollary}[theorem]{Corollary}
\newtheorem{prop}[theorem]{Proposition}

\newtheoremstyle{def}
{9pt}
{9pt}
{}
{}
{\bfseries}
{.}
{ }
{}
\theoremstyle{def}

\newtheorem{definition}[theorem]{Definition}
\newtheorem{remark}[theorem]{Remark}
\newtheorem{example}[theorem]{Example}

\newcommand{\R}{\mathbb{R}} 
\newcommand{\Z}{\mathbb{Z}} 
\newcommand{\N}{\mathbb{N}} 
\newcommand{\E}{\mathbb{E}} 
\newcommand{\PP}{\mathbb{P}} 

\renewcommand{\footnoterule}{%
	\kern -3.5pt
	\hrule width \textwidth height 1pt
	\kern 3.5pt
}

\makeatletter
\def\blfootnote{\xdef\@thefnmark{}\@footnotetext}
\makeatother

\title{Characterizations of non-normalized discrete probability distributions and their application in statistics}

\author{S. Betsch\\
Institute of Stochastics,\\ Karlsruhe Institute of Technology (KIT),\\ Germany.\\
\href{mailto:Steffen.Betsch@kit.edu}{Steffen.Betsch@kit.edu}\\
\And  B. Ebner\\
Institute of Stochastics,\\ Karlsruhe Institute of Technology (KIT),\\ Germany.\\
\href{mailto:Bruno.Ebner@kit.edu}{Bruno.Ebner@kit.edu}\\
\And F. Nestmann\\
Institute of Stochastics,\\ Karlsruhe Institute of Technology (KIT),\\ Germany.\\
\href{mailto:Franz.Nestmann2@kit.edu}{Franz.Nestmann2@kit.edu}\\
}

\begin{document}

\date{\today}
\maketitle

\blfootnote{\hspace{31mm}non-normalized models; Stein characterizations}

\begin{abstract}
	From the distributional characterizations that lie at the heart of Stein's method we derive explicit formulae for the mass functions of discrete probability laws that identify those distributions. These identities are applied to develop tools for the solution of statistical problems. Our characterizations, and hence the applications built on them, do not require any knowledge about normalization constants of the probability laws. To demonstrate that our statistical methods are sound, we provide comparative simulation studies for the testing of fit to the Poisson distribution and for parameter estimation of the negative binomial family when both parameters are unknown. We also consider the problem of parameter estimation for discrete exponential-polynomial models which generally are non-normalized.
\end{abstract}

\section{Introduction}
\label{SEC introduction}

In recent years, research in probability and statistics has witnessed the rise of Stein's method but also the emergence of methods to tackle the analysis and application of models which are based on non-normalized probability laws. In this work, we seek to apply findings from the research on Stein's method to contribute to the solution of testing and estimation problems involving non-normalized statistical distributions. We focus on the analysis of discrete probability laws, and how the theoretical results can be used to develop statistical methods. As such, we tie on to \cite{BE:2019:1} who provide similar tools for continuous distributions. A rather well-known approach to the problem of parameter estimation for non-normalized continuous probability distributions is the score matching technique due to \cite{H:2005, H:2007} \citep[see][for recent progress]{YDS:2019}. Another approach is known as noise contrastive estimation \citep[cf.][]{GH:2010}, but a number of 2019/20 papers indicate that the proposition and study of new tools remains an important issue, see \cite{MH:2019}, \cite{UKTM:2019}, and \cite{UMK:2019}.

The tool box which is now known as Stein's method goes back to the work of \cite{S:1972} \citep[see also][]{S:1986} who sought for an alternative proof of the central limit theorem that provides a bound on the rate of the convergence. This inherent feature made Stein's method popular. The idea is applied in all kinds of settings, as it allows to find bounds on distributional distances between sequences of probability laws and a limit distribution, and as it often applies in the absence of stochastic independence. The application of the method to discrete distributions goes back to \cite{Ch:1975}, who first derived corresponding results for the Poisson distribution, known as the Stein-Chen method. The method has since been extended to other discrete distributions, like the binomial distribution \citep[by][]{E:1991}, the geometric distribution \citep[by][]{P:1996}, the negative binomial distribution \citep[by][]{BP:1999}, discrete Gibbs measures \citep[by][]{ER:2008}, and others. The foundation of the method are characterization results for the underlying probability law. While for the first distributions in consideration specific identities were used or devised, general approaches have emerged that apply to many different distributions at once. In this context, we mention the generator approach of \cite{B:1988, B:1990} and \cite{G:1991} who use time-reversible Markov processes, where the stationary distribution is the probability law of interest, to characterize that law. On the other hand, a direct derivation of characterizations is possible, and a well-known class of such identities can be found under the name of 'density approach'. For the continuous case, first ideas on the density approach came from \cite{S:1986} and \cite{SDHR:2004}, and a more complete version is due to \cite{LS:2013:2}. The corresponding characterizations for discrete distributions are given by \cite{LS:2013}.

The contribution at hand is certainly not the first application of Stein's method in statistics. Indeed, similar problems in the context of non-normalized models are tackled with the use of so-called Stein discrepancies by the machine learning community, though many of them refer to the continuous setting. Let us mention some papers that explore these tools. Namely, \cite{CSG:2016}, \cite{LLJ:2016}, and \cite{YLRN:2018} consider the construction of tests of fit, \cite{GM:2015} build measures of sample quality, and \cite{BBDGM:2019} solve estimation problems for non-normalized models.

Our new work is based, in a strict sense, not on what is generally called Stein's method, but rather on the characterization identities we refer to above. More precisely, we take as a starting point the discrete density approach identity as provided by \cite{LS:2013}. To sketch the idea, consider a probability mass function $p$ on $\N_0$ as well as an $\N_0$-valued random variable $X$. Subject to few regularity conditions, $X$ is governed by $p$ if, and only if,
\begin{align*}
	\E \bigg[ \Delta^+ f(X) + \frac{\Delta^+ p(X)}{p(X)} \, f(X + 1) \bigg] = 0
\end{align*}
holds for a large enough class of test functions $f$. Hereby $\Delta^+$ denotes the forward difference operator. Our first contribution lies in proving that this characterization can essentially be restated as to $X$ being governed by $p$ if, and only if, the probability mass function $\rho_X$ of $X$ satisfies
\begin{align*}
	\rho_X(k)
	= \E \bigg[ - \frac{\Delta^+ p(X)}{p(X)} \, \mathds{1}\{ X \geq k \} \bigg], \quad k \in \N_0 .
\end{align*}
With regard to applications in statistics, this second identity is more accessible. We can, for instance, tackle the goodness-of-fit testing problem as follows. Assume we are to test whether a sample $X_1, \dots, X_n$ of $\N_0$-valued random variables follows one of the laws of a parametric family of distributions $\{ p_\vartheta : \vartheta \in \Theta \}$, where $\Theta$ denotes the parameter space. By the above characterization, if $X_1, \dots, X_n$ are governed by one of the $p_\vartheta$, then the difference between
\begin{align*}
	\widehat{\rho}_n(k)
	= \frac{1}{n} \sum_{j = 1}^n \mathds{1}\{ X_j = k \}
\end{align*}
and
\begin{align*}
	\frac{1}{n} \sum_{j = 1}^n - \frac{\Delta^+ p_{\widehat{\vartheta}_n}(X_j)}{p_{\widehat{\vartheta}_n}(X_j)} \, \mathds{1}\{ X_j \geq k \}
\end{align*}
ought to be small for each $k \in \N_0$. Here we denote by $\widehat{\vartheta}_n$ an estimator of $\vartheta$ based on $X_1, \dots, X_n$. Thus, in line with the idea of characterization based goodness-of-fit testing, our proposal is to use
\begin{align*}
	\sum_{k = 0}^\infty \bigg( \frac{1}{n} \sum_{j = 1}^n \mathds{1}\{ X_j = k \} + \frac{1}{n} \sum_{j = 1}^n \frac{\Delta^+ p_{\widehat{\vartheta}_n}(X_j)}{p_{\widehat{\vartheta}_n}(X_j)} \, \mathds{1}\{ X_j \geq k \} \bigg)^2
\end{align*}
as a test statistic for the hypothesis
\begin{align*}
	\mathbf{H_0} ~ : ~ \rho_{X_1} \in \{ p_\vartheta : \vartheta \in \Theta \} ,
\end{align*}
and to reject the hypothesis for large values of the statistic. Supposing that $X_1, \dots, X_n$ are governed by $p_{\vartheta_0}$ for some (unknown) $\vartheta_0 \in \Theta$, the very same heuristic leads us to propose
\begin{align*}
	\widehat{\vartheta}_n
	= \mbox{argmin}_{\vartheta \in \Theta} ~ \sum_{k = 0}^\infty \bigg( \frac{1}{n} \sum_{j = 1}^n \mathds{1}\{ X_j = k \} + \frac{1}{n} \sum_{j = 1}^n \frac{\Delta^+ p_{\vartheta}(X_j)}{p_{\vartheta}(X_j)} \, \mathds{1}\{ X_j \geq k \} \bigg)^2
\end{align*}
as an estimator for the unknown $\vartheta_0$. The paper at hand formalizes these ideas and puts them on firm mathematical ground. We also provide examples for the theoretical results as well as for the testing and estimation methods we propose. In Section \ref{SEC foundation} we introduce basic notation and recall the density approach identity. In Section \ref{SEC distri chara pmf} we prove the new characterization result as indicated above. Section \ref{SEC distri chara trafos of pmf} contains further characterizations based on transformations of the probability mass function such as distribution functions, characteristic functions, and generating functions. In Section \ref{SEC example distributions} we discuss examples. We then construct and study empirically, in Section \ref{SEC Poisson goodness-of-fit}, the test of fit for the Poisson distribution. In Section \ref{SEC negative binomial goodness-of-fit} a discrepancy measure as above leads to minimum distance estimators for the negative binomial distribution, which are put to a test in a simulation study. Section \ref{SEC discrete exponential-polynomial parameter estimation} deals with similar parameter estimators in the non-normalized class of discrete exponential-polynomial models.

\section{The foundation: Stein characterizations for discrete distributions}
\label{SEC foundation}

We denote by $p : \Z \to [0, 1]$ a probability mass function (pmf) defined on the integers. We assume that the support of $p$, that is, $\mathrm{spt}(p) = \big\{ k \in \Z : p(k) > 0 \big\}$ is connected, in the sense that
\begin{itemize}
	\item[(C1)] $\mathrm{spt}(p) = \{L, L + 1, \dots, R\}$, where $L, R \in \Z \cup \{ \pm \infty \}$, $L < R$ .
\end{itemize}
This prerequisite is quite usual in the context of Stein's method in the discrete setting. We further assume that
\begin{itemize}
	\item[(C2)] $\displaystyle \sup_{k \in \{ L, \dots, R - 1 \}} \bigg| \frac{\Delta^{+} p(k) \cdot \min\{ P(k), 1 - P(k) \}}{p(k) \, p(k + 1)} \bigg| < \infty$,
\end{itemize}
where $\Delta^+ f(k) = f(k + 1) - f(k)$ is the forward difference operator. Moreover, we denote by $P(k) = \sum_{\ell = L}^{k} p(\ell)$ the distribution function corresponding to $p$. Assumption (C2) is known from the continuous setting, see Lemma 13.1 of \cite{CGS:2011}. The supremum in (C2) runs from $L$ to $\infty$ whenever $R = \infty$. In what follows, we stick to the convention of empty sums being set to $0$.
\begin{definition} \label{DEF test functions}
	Let $p$ be a pmf that satisfies (C1) and (C2). We write $\mathcal{F}_p$ for the class of functions $f : \{ L, \dots, R \} \to \R$ such that
	\begin{itemize}
		\setlength{\itemindent}{-2.5mm}
		\item[($a$)] $\displaystyle \sum_{k = L}^R \big| \Delta^+ \big( p(k) \, f(k) \big) \big| < \infty \,$ and $\,\displaystyle \sum_{k = L}^R \Delta^+ \big( p(k) \, f(k) \big) = 0$, where we put $f(R + 1) = 0$ if $R < \infty$, as well as
		
		\item[($b$)] $\displaystyle \sup_{k \in \{ L, \dots, R \}} \big| \Delta^+ f(k) \big| < \infty$,\, and $\displaystyle \sup_{k \in \{ L, \dots, R \}} \bigg| \frac{\Delta^+ p(k)}{p(k)} \, f(k + 1) \bigg| < \infty$.
	\end{itemize}
\end{definition}
Conditions (C2) and ($b$) vanish completely whenever the support of $p$ is finite. We now state the characterization theorem that is known as Stein's density approach for discrete distributions. The proof is an easy adaptation of the proof of Theorem 2.1 from \cite{LS:2013} taking into account the different class of test functions. We give a full proof in Appendix \ref{APP SEC proof of density approach} to make it possible for the reader to comprehend how the assumptions come into play. Denote by $(\Omega, \mathcal{A}, \PP)$ the probability space which underlies all random quantities in this work.
\begin{theorem}[Discrete density approach] \label{THM density approach}
	Let $p$ be a pmf which satisfies (C1) and (C2), and let $X : \Omega \to \R$ be a random variable such that $\PP \big( X \in \mathrm{spt}(p) \big) > 0$. Then, $\PP\big( X = k \, | \, X \in \mathrm{spt}(p) \big) = p(k)$, $k \in \Z$, if, and only if,
	\begin{align*}
		\E \bigg[ \Delta^+ f(X) + \frac{\Delta^+ p(X)}{p(X)} \, f(X + 1) ~ \bigg| ~ X \in \mathrm{spt}(p) \bigg] = 0,
	\end{align*}
	for all $f \in \mathcal{F}_p$, where $\E[\cdot \, | \, \cdot]$ denotes the conditional expectation.
\end{theorem}
We use the abbreviation $X|p \sim p$ for $\PP\big( X = k \, | \, X \in \mathrm{spt}(p) \big) = p(k)$, $k \in \Z$. There exists a very similar result for continuous probability distributions. This continuous version existed first and was initiated by \cite{S:1986}. For the fully prepared statement we refer to \cite{LS:2011} and \cite{LS:2013:2}, and for further constructions of this type of Stein operators, see \cite{LRS:2017}.

\begin{remark} \label{RMK test functions}
	It follows from the proof of Theorem \ref{THM density approach} that, in the case $L > - \infty$, we may assume that $f(L) = 0$ for all $f \in \mathcal{F}_p$.
\end{remark}

\section{Distributional characterizations via the probability mass function}
\label{SEC distri chara pmf}

In this section, we derive explicit distributional characterizations via the probability mass function. The whole theory can be understood as a discrete version of the results from \cite{BE:2019:1} who established similar characterization identities for continuous probability laws starting with a continuous version of Theorem \ref{THM density approach} as stated by \cite{LS:2013:2}. We make the further assumption that the expectation of $p$ exists, that is,
\begin{itemize}
	\item[(C3)] $\E |Z| < \infty$, where $Z$ is a discrete random variable with pmf $p$.
\end{itemize}
It follows from (C3) that $\E \big| \Delta^+ p(Z) \cdot Z \, / \, p(Z) \big| < \infty$, and hence we also have $\E \big| \Delta^+ p(Z) \, / \, p(Z) \big| < \infty$. We note our first result, a proof of which is given in Appendix \ref{APP SEC proof of pmf chara}.
\begin{theorem} \label{THM chara pmf forward difference}
	Let $p$ be a pmf which satisfies (C1) -- (C3) with $L > - \infty$. Let $X : \Omega \to \R$ be a random variable with $\PP \big( X \in \mathrm{spt}(p) \big) > 0$ as well as
	\begin{align*}
		\E \bigg| \frac{\Delta^+ p(X)}{p(X)} \, X \cdot \mathds{1}\big\{ X \in \mathrm{spt}(p) \big\} \bigg| < \infty,
	\end{align*}
	and denote by $\rho_{X|p}(k) = \PP \big( X = k \, | \, X \in \mathrm{spt}(p) \big)$ the pmf of $X$ given $X \in \mathrm{spt}(p)$. Then, $X|p \sim p$ if, and only if,
	\begin{align*}
		\rho_{X|p}(k) = \E \bigg[ - \frac{\Delta^+ p(X)}{p(X)} \, \mathds{1}\{ X \geq k \} ~ \bigg| ~ X \in \mathrm{spt}(p) \bigg] , \quad k \in \Z, k \geq L.
	\end{align*}
\end{theorem}
Notice that the integrability assumption on $X$ implies the existence of the (conditional) expectation that appears in the theorem. Even in stating Theorem \ref{THM chara pmf forward difference} the ordering of the integers is essential. However, if $p$ is an admissible probability mass function on some arbitrary countable set $\mathbb{S}$ (where $\mathbb{S}$ is endowed with the power set as a $\sigma$-field), there exists a bijection $\iota : \mathbb{S} \to \mathbb{N}_0$, which corresponds to imposing an order on the space $\mathbb{S}$, and Theorem \ref{THM chara pmf forward difference} can be applied. This leads to the following corollary which allows the handling of more general state spaces.
\begin{corollary} \label{COR chara pmf general state space}
	Let $\mathbb{S}$ be a countable set and $p : \mathbb{S} \to [0, 1]$ such that $\sum_{s \in \mathbb{S}} p(s) = 1$. Let $\iota : \mathbb{S} \to \{ L, \dots, R \}$, with $L > - \infty$, be a bijection so that $\widetilde{p} = p \circ \iota^{-1}$ satisfies (C1) -- (C3). Assume that $X : \Omega \to \mathbb{S}$ is a random variable with $\PP\big( X \in \mathrm{spt}(p) \big) > 0$, and
	\begin{align*}
		\E \bigg| \frac{(\Delta^+ \widetilde{p})\big( \iota(X) \big)}{p(X)} \, \iota(X) \cdot \mathds{1}\big\{ X \in \mathrm{spt}(p) \big\} \bigg| < \infty .
	\end{align*}
	Then, $X|p \sim p$ if, and only if,
	\begin{align*}
		\rho_{X|p}(k) = \E \bigg[ - \frac{(\Delta^+ \widetilde{p})\big( \iota(X) \big)}{p(X)} \, \mathds{1}\big\{ \iota(X) \geq \iota(s) \big\} ~ \bigg| ~ X \in \mathrm{spt}(p) \bigg] , \quad s \in \mathbb{S} .
	\end{align*}
\end{corollary}
Any such ordering on $\mathbb{S}$ gives a characterization result, and if $X \sim p$, the (conditional) expectation is the same for every ordering (as $\rho_{X|p}$ does not depend on $\iota$). However, if one intends to use the converse of the characterization (with general $X$), the calculation of the expectation depends on the ordering, so in practice the question of choosing an efficient ordering arises. Finding an order such that the condition (C1) -- (C3) are satisfied is a non-trivial endeavor. We give one example of choosing an order such that a pmf with a support that is not bounded from below can be considered. To state the result, we first need to recall that the backward difference operator $\Delta^-$ is defined by $\Delta^- f(k) = f(k) - f(k - 1)$.
\begin{corollary} \label{COR chara pmf backward difference}
	Let $p$ be a pmf on $\{ L, \dots, R \}$ which satisfies (C1) -- (C3) with $R < \infty$. Let $X : \Omega \to \R$ be a random variable which satisfies $\PP \big( X \in \mathrm{spt}(p) \big) > 0$ as well as
	\begin{align*}
		\E \bigg| \frac{\Delta^- p(X)}{p(X)} \, X \cdot \mathds{1}\big\{ X \in \mathrm{spt}(p) \big\} \bigg| < \infty.
	\end{align*}
	Then, $X|p \sim p$ if, and only if,
	\begin{align*}
		\rho_{X|p}(k) = \E \bigg[ \frac{\Delta^- p(X)}{p(X)} \, \mathds{1}\{ X \leq k \} ~ \bigg| ~ X \in \mathrm{spt}(p) \bigg], \quad k \in \Z, k \leq R.
	\end{align*}
\end{corollary}
The result follows from Corollary \ref{COR chara pmf general state space} upon choosing $\iota : \{ L, \dots, R \} \to \{ -R, \dots, -L \}$, $\iota(k) = -k$, and observing that
\begin{align*}
	(\Delta^+ \widetilde{p})\big( \iota(X) \big)
	= (\Delta^+ \widetilde{p})(-X)
	= p(X - 1) - p(X)
	= - \Delta^- p(X) .
\end{align*}
Note that Corollary \ref{COR chara pmf backward difference} can also be obtained via a different path. With few technical changes in Definition \ref{DEF test functions} and Appendix \ref{APP SEC proof of density approach}, a $\Delta^-$-version of Theorem \ref{THM density approach} can be formulated \cite[see also][]{LS:2013}. Using this result and an adaptation of the proof of Theorem \ref{THM chara pmf forward difference} yields another proof of Corollary \ref{COR chara pmf backward difference}.

\begin{remark} \label{RMK chara theorem for pmf}
	 Whenever $X$ is assumed a priori to take values in $\{ L, \dots, R \}$, the conditioning on $X \in \mathrm{spt}(p)$ can be omitted, and when $- \infty < L < R < \infty$, the integrability condition on $X$ is trivially satisfied. As for the regularity assumptions (C1) -- (C3), notice that, by Corollary \ref{COR chara pmf general state space}, (C1) is mostly an issue of notation but no hard restriction. Whenever we deal with discrete distributions that have finite support, conditions (C2) and (C3) are trivially satisfied. In case of an infinite support, assumption (C3) is easy to interpret. It is stated to guarantee that the statement of Theorem \ref{THM chara pmf forward difference} is consistent, as it ensures that a random variable $Z \sim p$ satisfies the integrability condition on $X$. A drawback in terms of the assumptions is that we cannot give a general treatment of (C2), and that this condition can sometimes be difficult to check for a given distribution. A similar condition (with identical problems) is required by \cite{BE:2019:1} in the continuous setting. If $L > - \infty$ and $R = \infty$ then (C2) holds if, and only if,
	\begin{align} \label{equiv. condition for C2}
		\limsup_{k \, \to \, \infty} \bigg| \frac{\Delta^+p(k) \cdot (1 - P(k))}{p(k) \, p(k + 1)} \bigg| < \infty .
	\end{align}
	Similar thoughts apply to other choices for $L$ and $R$, but this does not solve the problem in general. However, the Stolz-Ces\'{a}ro theorem \citep[see Theorem 2.7.1 of][]{CN:2014} provides a useful tool for checking the condition in practice, see Example \ref{EXA exponential polynomial}.
\end{remark}

\begin{remark}[Non-normalized models] \label{RMK non-normalized models}
	As we have explained in the introduction, many statistical models, primarily in machine learning and physics, are too complex for the normalization constant of the distribution to be calculable. As estimation and testing procedures (e.g. the maximum likelihood estimator) normally rely on some knowledge about this constant, they may not be applicable to such models. Thus, we want to emphasize that our explicit characterizations do not need any knowledge about the normalization constants, and neither do any of the characterizations and statistical applications presented in subsequent sections.
\end{remark}

\section{Characterizations via transformations of the probability mass function}
\label{SEC distri chara trafos of pmf}
We also obtain characterizing formulae for transformations of the pmf, like distribution functions, characteristic functions, and probability generating functions. We focus on the identities derived for mass functions on the integers but more general spaces can be treated in the lines of Corollary \ref{COR chara pmf general state space}.
\begin{prop}[Distribution functions] \label{PROP chara distribution functions}
	Let $p$ be a pmf which satisfies (C1) -- (C3) with $L > - \infty$. Let $X : \Omega \to \R$ be a random variable with $\PP \big( X \in \mathrm{spt}(p) \big) > 0$ as well as
	\begin{align} \label{integrability assumption on X}
		\E \bigg| \frac{\Delta^+ p(X)}{p(X)} \, X \cdot \mathds{1}\big\{ X \in \mathrm{spt}(p) \big\} \bigg| < \infty,
	\end{align}
	and further denote by $F_{X|p}(k) = \PP \big( X \leq k \, | \, X \in \mathrm{spt}(p) \big)$ the distribution function of $X$ given $X \in \mathrm{spt}(p)$. Then, $X|p \sim p$ if, and only if,
	\begin{align*}
		F_{X|p}(k) = \E \bigg[ - \frac{\Delta^+ p(X)}{p(X)} \big( \min\{ X, k \} - L + 1 \big) ~ \bigg| ~ X \in \mathrm{spt}(p) \bigg] , \quad k \in \Z, k \geq L.
	\end{align*}
\end{prop}
The proof of this proposition is provided in Appendix \ref{APP SEC proof of distribution function chara}. We continue to give another characterization based on the characteristic function. The proof (see Appendix \ref{APP SEC proof of characteristic function chara}) features the inversion formula for characteristic functions. In the continuous setting, the inversion formula requires an integrability condition that is not needed in the discrete setting. If one finds a way to handle this integrability condition, similar identities are conceivable for the continuous setting of \cite{BE:2019:1}.
\begin{prop}[Characteristic functions] \label{PROP chara characteristic functions}
	Assume that $p$ is a pmf which satisfies (C1) -- (C3) with $L > - \infty$. Take $X : \Omega \to \R$ to be a random variable with $\PP \big( X \in \mathrm{spt}(p) \big) > 0$ such that assumption (\ref{integrability assumption on X}) holds. Denote by $\varphi_{X|p}(t) = \E \big[ e^{i t X} \, \big| \, X \in \mathrm{spt}(p) \big]$, $t \in \R$, the characteristic function of $X$ given $X \in \mathrm{spt}(p)$ (where $i$ is the complex unit). Then, $X|p \sim p$ if, and only if,
	\begin{align*}
		\varphi_{X|p}(t)
		= \E \bigg[ - \frac{\Delta^+ p(X)}{p(X)} \cdot \frac{e^{i t L} - e^{i t (X + 1)}}{1 - e^{i t}} ~ \bigg| ~ X \in \mathrm{spt}(p) \bigg] , \quad t \in \R.
	\end{align*}
\end{prop}


We conclude this section on transformations of the pmf with a characterization via the probability generating function, thus specializing to the case $\mathrm{spt}(p) = \N_0$. A proof is given in Appendix \ref{APP SEC proof of probability generating function chara}.
\begin{prop}[Generating functions] \label{PROP chara probability generating functions}
	Let $p$ be a pmf that satisfies (C1) -- (C3) with $\mathrm{spt}(p) = \N_0$. Let $X : \Omega \to \N_0$ be a discrete random variable such that
	\begin{align*}
		\E \bigg| \frac{\Delta^+ p(X)}{p(X)} \, X \bigg| < \infty.
	\end{align*}
	Denote by $G_{X}(s) = \E [s^X]$, $s \in [0, 1)$, the (probability) generating function of $X$. Then, $X \sim p$ if, and only if,
	\begin{align*}
		G_{X}(s)
		= \E \bigg[ - \frac{\Delta^+ p(X)}{p(X)} \cdot \frac{1 - s^{X + 1}}{1 - s} \bigg] , \quad s \in [0, 1).
	\end{align*}
\end{prop}

\section{Examples}
\label{SEC example distributions}

In this section, we provide examples that fit into our framework. For each distribution we indicate why (C1) -- (C3) hold and we explicitly state the characterization via Theorem \ref{THM chara pmf forward difference}, though sometimes in the unconditioned formulation. The characterizations via Propositions \ref{PROP chara distribution functions}, \ref{PROP chara characteristic functions}, and \ref{PROP chara probability generating functions} are not stated explicitly. We start by giving three infinite support examples which are subject of our statistical applications in the subsequent sections. More precisely, we discuss the Poisson and the negative binomial distribution as well as a discrete version of the exponential-polynomial model.
\begin{example}[Poisson distribution] \label{EXA Poisson}
	The mass function of the Poisson distribution is given as $p(k) = \lambda^k \cdot e^{-\lambda} \, / \, k!$, $k \in \N_0$, for some rate parameter $\lambda > 0$. In this case, we obtain
	\begin{align*}
	\frac{\Delta^+ p(k)}{p(k)}
	= \frac{\lambda}{k + 1} - 1, \quad k \in \N_0.
	\end{align*}
	Conditions (C1) and (C3) are obviously true. To see that (C2) holds, note that whenever $\tfrac{\lambda}{k + 2} < 1$, we have
	\begin{align*}
	\bigg| \frac{\Delta^{+} p(k) \cdot (1 - P(k))}{p(k) \, p(k + 1)} \bigg|
	= \bigg| \frac{\lambda}{k + 1} - 1 \bigg| \sum_{\ell = 1}^\infty \frac{\lambda^{\ell - 1}}{(\ell + k)!} \, (k + 1)!
	&\leq \bigg| \frac{\lambda}{k + 1} - 1 \bigg| \sum_{\ell = 0}^\infty \bigg( \frac{\lambda}{k + 2} \bigg)^\ell \\
	&= \bigg| \frac{\lambda}{k + 1} - 1 \bigg| \cdot \frac{1}{1 - \tfrac{\lambda}{k + 2}} ,
	\end{align*}
	and therefore (\ref{equiv. condition for C2}) holds which yields (C2). Theorem \ref{THM chara pmf forward difference} implies that a random variable $X : \Omega \to \N_0$ with $\E X < \infty$ has the Poisson distribution with parameter $\lambda$ if, and only if,
	\begin{align*}
	\rho_X(k)
	= \E \bigg[ \bigg( 1 - \frac{\lambda}{X + 1} \bigg) \mathds{1}\{ X \geq k \} \bigg], \quad k \in \N_0.
	\end{align*}
\end{example}

\begin{example}[Negative binomial distribution] \label{EXA negative binomial}
	Let $p(k) = {k + r - 1 \choose k} (1 - q)^k \, q^r$, $k \in \N_0$, be the probability mass function of the negative binomial distribution with parameters $r > 0$ and $q \in (0, 1)$. An important special case arises for $r = 1$, where the negative binomial distribution reduces to the geometric distribution. These laws are frequently used in the analysis of arrival times. We have
	\begin{align*}
	\frac{\Delta^+ p(k)}{p(k)}
	= \frac{r + k}{k + 1} \, (1 - q) - 1, \quad k \in \N_0.
	\end{align*}
	Condition (C1) is trivially satisfied, and (C3) is easily verified. We prove (\ref{equiv. condition for C2}) to show that (C2) is satisfied. To this end, observe that
	\begin{align*}
	\bigg| \frac{\Delta^{+} p(k) \cdot (1 - P(k))}{p(k) \, p(k + 1)} \bigg|
	&= \bigg| \frac{r + k}{k + 1} \, (1 - q) - 1 \bigg| \sum_{\ell = 0}^\infty \frac{(\ell + k + r)! \cdot (k + 1)!}{(\ell + k + 1)! \cdot (k + r)!} \, (1 - q)^\ell \\
	&= \bigg| \frac{r + k}{k + 1} \, (1 - q) - 1 \bigg| \sum_{\ell = 0}^\infty \frac{(r + k + \ell) \cdot (r + k + \ell - 1) \cdots (r + k + 1)}{(k + \ell + 1) \cdot (k + \ell) \cdots (k + 2)} \, (1 - q)^\ell.
	\end{align*}
	If $r \leq 1$, the sum on the right hand-side is bounded by $\sum_{\ell = 0}^\infty (1 - q)^\ell = \tfrac{1}{q}$. If $r > 1$, let $k$ be large enough so that $2 \cdot (r - 1) \, / \, (k + 2) < q \, / \, (1 - q)$, and observe that
	\begin{align*}
	\sum_{\ell = 0}^\infty \frac{(r + k + \ell) \cdots (r + k + 1)}{(k + \ell + 1) \cdots (k + 2)} \, (1 - q)^\ell
	&= \sum_{\ell = 0}^\infty \bigg( 1 + \frac{r - 1}{k + \ell + 1} \bigg) \cdot \bigg( 1 + \frac{r - 1}{k + \ell} \bigg) \cdots \bigg( 1 + \frac{r - 1}{k + 2} \bigg) \cdot (1 - q)^\ell \\
	&\leq \sum_{\ell = 0}^\infty \bigg( 1 + \frac{1}{2} \cdot \frac{q}{1 - q} \bigg)^\ell \, (1 - q)^\ell = \sum_{\ell = 0}^\infty \bigg(1 - \frac{q}{2} \bigg)^\ell
	= \frac{2}{q} ,
	\end{align*}
	where the products in the sum are empty (hence equal to $1$) for $\ell = 0$. In any case, (\ref{equiv. condition for C2}) follows, so (C2) is valid. Theorem \ref{THM chara pmf forward difference} states that a discrete random variable $X : \Omega \to \N_0$ with $\E X < \infty$ follows the negative binomial law with parameters $r$ and $q$ if, and only if,
	\begin{align*}
	\rho_X(k)
	= \E \bigg[ \bigg( 1 - \frac{r + X}{X + 1} \, (1 - q) \bigg) \mathds{1}\{ X \geq k \} \bigg], \quad k \in \N_0.
	\end{align*}
	Note that the statement by \cite{JKK:1993} (on p. 223) that "only a few characterizations have been obtained for the negative binomial distribution" appears to still hold true. For one recent characterization related to Stein's method, we refer to \cite{AH:2019}.
\end{example}

\begin{example}[Exponential-polynomial models] \label{EXA exponential polynomial}
	We consider the following discrete exponential-polynomial parametric model given through
	\begin{align*}
		p_\vartheta(k)
		= C(\vartheta)^{-1} \exp\big( \vartheta_1 k + \dotso + \vartheta_d k^d \big), \quad k \in \N,
	\end{align*}
	where
	\begin{align*}
		C(\vartheta)
		= \sum_{k = 1}^\infty \exp\big( \vartheta_1 k + \dotso + \vartheta_d k^d \big) ,
	\end{align*}
	and $\vartheta = (\vartheta_1, \dots, \vartheta_d) \in \R^{d - 1} \times (- \infty, 0)$. This corresponds to a discrete exponential family in the canonical form with the sufficient statistic containing monomials up to order $d \in \N$, with $d \geq 2$. Clearly condition (C1) is satisfied and the restriction $\vartheta_d < 0$ ensures that (C3) holds for every $\vartheta$ as well as that $C(\vartheta) < \infty$.
	We have
	\begin{align*}
		\frac{p_\vartheta(k + 1)}{p_\vartheta(k)}
		= \exp\Big( \vartheta_1 + \vartheta_2 \big( (k + 1)^2 - k^2 \big) + \dotso + \vartheta_d \big( (k + 1)^d - k^d \big) \Big)
		\longrightarrow 0 ,
	\end{align*}
	as $k \to \infty$, and the Stolz-Ces\'{a}ro theorem \cite[Theorem 2.7.1 of][]{CN:2014} yields
	\begin{align*}
		\lim_{k \, \to \, \infty} \frac{p_\vartheta(k + 1)}{1 - P_\vartheta(k)}
		= - \lim_{k \, \to \, \infty} \frac{p_\vartheta(k + 2)}{p_\vartheta(k + 1)} + 1
		= 1 .
	\end{align*}
	Consequently, we obtain
	\begin{align*}
		\limsup_{k \, \to \, \infty} \bigg| \frac{\Delta^+ p_\vartheta(k) \cdot (1 - P_\vartheta(k))}{p_\vartheta(k) \, p_\vartheta(k + 1)} \bigg|
		\leq \limsup_{k \, \to \, \infty} \bigg| \frac{p_\vartheta(k + 1)}{p_\vartheta(k)} - 1 \bigg| \cdot \bigg| \frac{1 - P_\vartheta(k)}{p_\vartheta(k + 1)} \bigg|
		= 1
	\end{align*}
	for every $\vartheta \in \R^{d - 1} \times (- \infty, 0)$, so (C2) holds. Finally observe that, since $p_\vartheta(k + 1) \, / \, p_\vartheta(k) < 1$ for all but finitely many $k \in \N$, an $\N$-valued random variable with $\E X < \infty$ also satisfies
	\begin{align*}
		\E \bigg| \frac{\Delta^+ p_\vartheta(X)}{p_\vartheta(X)} \, X \bigg|
		< \infty .
	\end{align*}
	Theorem \ref{THM chara pmf forward difference} yields that a random variable $X : \Omega \to \N$ with $\E X < \infty$ has the pmf $p_\vartheta$ if, and only if,
	\begin{align*}
		\rho_X(k)
		= \E \bigg[ \bigg( - \exp\Big( \vartheta_1 + \vartheta_2 \big( (X + 1)^2 - X^2 \big) + \dotso + \vartheta_d \big( (X + 1)^d - X^d \big) \Big) + 1 \bigg) \mathds{1}\{ X \geq k \} \bigg], \quad k \in \N.
	\end{align*}
	In Section \ref{SEC discrete exponential-polynomial parameter estimation} we use this characterization to construct an estimation method for this type of parametric model, focusing on a two-parameter case where $d = 3$ and $\vartheta_2 = 0$ fixed.
\end{example}

We now take a look at the resulting characterization for the uniform and binomial distribution.
\begin{example}[Uniform distribution] \label{EXA uniform}
	Assume that $p$ is given through $p(k) = 1 / m$, for $k = 1, \dots, m$, $m \geq 2$. Then, (C1) -- (C3) are obviously satisfied (recall Remark \ref{RMK chara theorem for pmf}), and we have
	\begin{align*}
		\frac{\Delta^+ p(k)}{p(k)} = 0, \quad \text{for } k = 1, \dots, m - 1, \quad\quad \text{and} \quad \quad \frac{\Delta^+ p(m)}{p(m)} = -1.
	\end{align*}
	By Theorem \ref{THM chara pmf forward difference}, a random variable $X : \Omega \to \R$ with $\PP \big( X \in \{ 1, \dots, m \} \big) > 0$ satisfies $X|p \sim p$ if, and only if,
	\begin{align*}
		\rho_{X|p}(k) = \frac{\PP (X = m)}{\PP \big( X \in \{ 1, \dots, m \} \big)}, \quad k = 1, \dots, m,
	\end{align*}
	that is, if $\rho_{X|p}(k) = 1 / m$, for $k = 1, \dots, m$. This result is easily derived by a direct argument, so for the uniform distribution, our characterization contains no new information. A similar behavior was observed in the continuous setting by \cite{BE:2019:1}. Note that Corollary \ref{COR chara pmf backward difference} leads to the very same characterization as Theorem \ref{THM chara pmf forward difference}.
\end{example}

\begin{example}[Binomial distribution] \label{EXA binomial}
	Let $p(k) = {m \choose k}  q^k (1 - q)^{m - k}$, for $k = 0, \dots, m$, $m \geq 1$, and some fixed $q \in (0, 1)$. Then, we have
	\begin{align*}
		\frac{\Delta^+ p(k)}{p(k)} = \frac{q}{1 - q} \cdot \frac{m - k}{k + 1} - 1, \quad \text{for } k = 0, \dots, m - 1, \quad\quad \text{and} \quad\quad \frac{\Delta^+ p(m)}{p(m)} = -1.
	\end{align*}
	By Theorem \ref{THM chara pmf forward difference}, a random variable $X : \Omega \to \R$ with $\PP \big( X \in \{ 0, \dots, m \} \big) > 0$ satisfies $X|p \sim p$ if, and only if,
	\begin{align*}
		\rho_{X|p}(k)
		= \frac{\sum_{\ell = k}^{m - 1} \big( 1 - \frac{q}{1 - q} \cdot \frac{m - \ell}{\ell + 1} \big) \, \PP(X = \ell) + \PP(X = m)}{\PP \big( X \in \{ 0, \dots, m \} \big)}, \quad k = 0, \dots, m.
	\end{align*}
	Moreover, we have
	\begin{align*}
		\frac{\Delta^- p(k)}{p(k)} = 1 - \frac{1 - q}{q} \cdot \frac{k}{m - k + 1}, \quad \text{for } k = 1, \dots, m, \quad\quad \text{and} \quad\quad \frac{\Delta^- p(0)}{p(0)} = 1,
	\end{align*}
	so Corollary \ref{COR chara pmf backward difference} yields that $X|p \sim p$ if, and only if,
	\begin{align*}
		\rho_{X|p}(k)
		= \frac{\sum_{\ell = 1}^{k} \big( 1 - \frac{1 - q}{q} \cdot \frac{\ell}{m - \ell + 1} \big) \, \PP(X = \ell) + \PP(X = 0)}{\PP \big( X \in \{ 0, \dots, m \} \big)}, \quad k = 0, \dots, m.
	\end{align*}
	Thus the example of the binomial distribution shows that Theorem \ref{THM chara pmf forward difference} and Corollary \ref{COR chara pmf backward difference} do not always lead to the same identity in cases where both are applicable.
\end{example}

We conclude this section on examples by showing that discrete Gibbs measures, which describe physical systems with countably many states, also fall into our framework.
\begin{example}[Gibbs (or Boltzmann) distribution] \label{EXA Gibbs distribution}
	We distinguish in our discussion between finite and infinite support.
	\begin{itemize}
		\item Assume that a given system can have $S \in \N$ states, $S \geq 2$, and let $V : \{1, \dots, S\} \to \R \cup \{ \infty \}$ be a map (called the energy function) which assigns each state its corresponding energy. Another map $N : \{ 1, \dots S \} \to \N$ assigns to each state the number of particles the system has in the given state. Let $\mu \in \R$ (the chemical potential), $T > 0$ (the temperature) and denote by $\kappa$ the Boltzmann constant ($\approx 1.380649 \cdot 10^{-23}$ joule per kelvin). The Gibbs distribution is given through
		\begin{align*}
			p(k) = \frac{1}{\Xi} \, \exp\bigg( \frac{\mu \, N(k) - V(k)}{\kappa \, T} \bigg), \quad k \in \{ 1, \dots, S \},
		\end{align*}
		where
		\begin{align*}
			\Xi = \Xi(\mu, T, N, V)
			= \sum_{\ell = 1}^{S} \exp\bigg( \frac{\mu \, N(\ell) - V(\ell)}{\kappa \, T} \bigg),
		\end{align*}
		and where we assume that $V(k) < \infty$ for at least one $k \in \{ 1, \dots S \}$ (which ensures $\Xi > 0$). In this setting with finitely many states, (C1) -- (C3) are trivially satisfied. We obtain $\Delta^+ p(S) \, / \, p(S) = -1$, and
		\begin{align*}
			\frac{\Delta^+ p(k)}{p(k)}
			= \exp\bigg( \frac{V(k) - V(k + 1) + \mu \big( N(k + 1) - N(k) \big)}{\kappa \, T} \bigg) - 1, \quad k = 1, \dots, S - 1.
		\end{align*}
		Theorem \ref{THM chara pmf forward difference} yields that a random variable $X :\Omega \to \{ 1, \dots, S \}$ follows the Gibbs distribution as above if, and only if, for all $k = 1, \dots, S$,
		\begin{align*}
			\rho_X(k)
			= \sum_{\ell = k}^{S - 1} \bigg( 1 - \exp\bigg( \frac{V(\ell) - V(\ell + 1) + \mu \big( N(\ell + 1) - N(\ell) \big)}{\kappa \, T} \bigg) \bigg) \PP\big( X = \ell \big) + \PP\big( X = S \big).
		\end{align*}
		
		\item The Gibbs distribution immediately generalizes to the case where $S = \infty$, that is, a system with infinitely many possible states. In this general setting however, one has to make further assumptions on $\mu$, $T$, $N$, and $V$ to ensure that $\Xi < \infty$ and that (C2) and (C3) hold. One set of assumptions that ensures $\Xi < \infty$ and (C3) is that the number of particles is fixed $N \equiv \overline{N} \in \N$ and that the probability of the system being in one of the states with higher index decreases sufficiently fast, or equivalently, that the energy grows fast enough. More precisely, it is sufficient to assume that there exists some $k_0 \in \N$ such that, for all $k \geq k_0$, we have $V(k) \geq c \cdot k^\alpha$, for some $c > 0$ and $\alpha \geq 1$. In order for (C2) to hold, we could additionally assume that there exists some $k_1 \in \N$ and $c^\prime > 0$ such that, for all $k \geq k_1$,
		\begin{align*}
			V(k) - V(k + n) \leq - c^\prime \, n, \quad n \in \N_0,
		\end{align*}
		as this implies
		\begin{align*}
			\bigg| \frac{\Delta^{+} p(k) \cdot \min\{ P(k), 1 - P(k) \}}{p(k) \, p(k + 1)} \bigg|
			\leq 2 \sum_{n = 0}^\infty \bigg( e^{- \tfrac{c^\prime}{\kappa \, T}} \bigg)^n
			< \infty, \quad k \geq k_1.
		\end{align*}
		One choice of $V$ (to satisfy all of the conditions) is thus $V(k) = C \cdot k$ for all $k$ larger than some fixed $k^* \in \N$ and some $C > 0$. The characterization via Theorem \ref{THM chara pmf forward difference} in the case of infinitely many possible states is similar as in the finite case: We have, for any $k \in \N$,
		\begin{align*}
			\frac{\Delta^+ p(k)}{p(k)}
			= \exp\bigg( \frac{V(k) - V(k + 1)}{\kappa \, T} \bigg) - 1.
		\end{align*}
		A random variable $X : \Omega \to \N$ with $\E X < \infty$ follows the (infinite states) Gibbs distribution if, and only if,
		\begin{align*}
			\rho_X(k)
			= \E \bigg[ \bigg( 1 - \exp\bigg( \frac{V(X) - V(X + 1)}{\kappa \, T} \bigg) \bigg) \mathds{1}\{X \geq k\} \bigg] , \quad k \in \N.
		\end{align*}
	\end{itemize}
	As is indicated by the names that certain quantities in the above display of the Gibbs distribution carry, the model appears in statistical mechanics, see the reprint, \cite{G:2010}, of Josiah Willard Gibbs' work from 1902, or virtually any textbook on statistical mechanics. It also plays an important role in image analysis and processing, see \cite{L:2009}. A crucial observation about our characterizations is that the partition function $\Xi$ vanishes completely.
\end{example}

\section{Goodness-of-fit testing for the Poisson distribution}
\label{SEC Poisson goodness-of-fit}

A first application of the characterization results from the previous sections is the construction of a test of fit for the Poisson distribution. Given a sample of $\N_0$-valued independent identically distributed (i.i.d.) random variables, the problem is to test the composite hypothesis that the sample comes from some Poisson distribution $\mathrm{Po}(\lambda)$ with an unknown rate parameter $\lambda > 0$, that is,
\begin{align*}
	\mathbf{H_0} ~ : ~ \mathbb{P}^{X_1} \in \big\{ \mathrm{Po}(\lambda) : \lambda > 0 \big\}.
\end{align*}
This is a classical statistical problem, well studied in the literature. Apart from Pearson's $\chi^2$ test, see \cite{K:2013} for recent developments, the hitherto proposed tests are based on the (conditional) empirical distribution function, see \cite{BB:2019,GH:2000,H:1996,F:2012}, the empirical probability generating function, see \cite{BH:1992,PW:2020,RO:1999}, on the integrated distribution function, see \cite{K:1999}, on a characterization by mean distance, see \cite{SR:2004}, on quadratic forms of score vectors, see \cite{I:2019}, on Charlier polynomials, see \cite{LW:2017}, on conditional probabilities ratio, see \cite{BB:2019}, and on relating first- and second-order moments, see \cite{KLP:1998}. For a survey of classical procedures and a comparative simulation study see \cite{GH:2000}. However, the construction of new and powerful methods is still of relevance: As \cite{N:2017} stated on p.4 of his contribution that "[...] one should keep in mind that any hypothesis has to be tested with several possible criteria. The point of the matter is that with absolute confidence we can only reject it, while each new test which fails to reject the null-hypothesis gradually brings the statistician closer to the perception that this hypothesis is true".

The idea of our new method is to estimate the two quantities that appear in the characterization via Theorem \ref{THM chara pmf forward difference} as given in Example \ref{EXA Poisson}, and to compare these empirical quantities. Based on the sample $X_1, \dots, X_n$, let
\begin{align*}
	\widehat{e}_n(k) = \frac{1}{n} \sum_{j = 1}^{n} \bigg( 1 - \frac{\widehat{\lambda}_n}{X_j + 1} \bigg) \mathds{1}\{ X_j \geq k \}, \quad k \in \N_0,
\end{align*}
be an estimator of the expectation that arises in the characterization, where $\widehat{\lambda}_n = n^{-1} \sum_{j = 1}^n X_j$ is a consistent estimator of the rate parameter. Also consider the empirical probability mass function,
\begin{align*}
	\widehat{\rho}_n(k) = \frac{1}n \sum_{j = 1}^n \mathds{1}\{ X_j = k \}, \quad k \in \N_0,
\end{align*}
as an estimator of $\rho_{X_1}$. By Theorem \ref{THM chara pmf forward difference} (see Example \ref{EXA Poisson}), if the sample $X_1, \dots, X_n$ comes from a Poisson distribution, the absolute difference between $\widehat{e}_n(k)$ and $\widehat{\rho}_n(k)$ ought to be small for every $k \in \N_0$. On the other hand, if the sample does not come from the Poisson law, we expect their absolute difference to be large. Based on this heuristic, we suggest to use as a test statistic the squared difference of $\widehat{e}_n$ and $\widehat{\rho}_n$ summed over $k \in \N_0$, that is,
\begin{align*}
	T_n^{Po}
	= \sum_{k = 0}^\infty \big( \widehat{e}_n(k) - \widehat{\rho}_n(k) \big)^2 ,
\end{align*}
and to reject the Poisson hypothesis $\mathbf{H_0}$ for large values of $T_n^{Po}$. Note that we do not need to introduce any weight functions to make the infinite sum in the definition of $T_n^{Po}$ converge, and observe that we choose the squared distance to obtain a finite double sum representation for $T_n^{Po}$, namely
\begin{align*}
	T_n^{Po}
	&= \frac{1}{n^2} \sum_{j, \, \ell = 1}^n \bigg[ \bigg( 1 - \frac{\widehat{\lambda}_n}{X_j + 1} \bigg) \Big( X_\ell - 1 - \widehat{\lambda}_n \Big) \, \mathds{1}\{ X_j \geq X_\ell \} \\
	&\qquad\qquad\qquad + \Big( X_j + 1 - \widehat{\lambda}_n \Big) \bigg( 1 - \frac{\widehat{\lambda}_n}{X_\ell + 1} \bigg) \mathds{1}\{ X_j < X_\ell \} + \mathds{1}\{ X_j = X_\ell \} \bigg] ,
\end{align*}
which is a numerically stable representation, hence easily implemented in a computer. The calculation of $T_n^{Po}$ involves only straight forward algebra and consists, mainly, of writing the squared difference of $\widehat{e}_n(k)$ and $\widehat{\rho}_n(k)$ as a double sum, multiplying the corresponding terms, and solving the sum over $k$ of the indicator functions.
\begin{remark}
	The major advantage in using Theorem \ref{THM chara pmf forward difference} to construct the test is that the empirical quantities are integrable with respect to the counting measure $\sum_{k = 0}^{\infty} \delta_k$. Of course, the Propositions \ref{PROP chara distribution functions}, \ref{PROP chara characteristic functions}, and \ref{PROP chara probability generating functions} provide similar heuristics for the construction of goodness-of-fit tests [see also \cite{BE:2019:2}, \cite{ABEV:2019}, and \cite{BE:2020} for the continuous setting], but the quantities are not necessarily integrable/summable and thus require the introduction of some weight function. What is more, it seems difficult to obtain explicit formulae as in the previous equation for $T_n^{Po}$, so the routines would rely on numerical integration which is computationally costly. We must therefore leave it as a problem of further research to employ these characterizations via distribution, characteristic, or generating functions in the construction of goodness-of-fit tests.
\end{remark}
As a proof of concept, we carry out a simulation study in order to compare our new test of poissonity with established procedures. All simulations are performed in the statistical computing environment \texttt{R}, see \cite{r20}.  We consider the sample size $n = 50$ and the nominal level of significance is set to $0.05$. Based on the methodology for asymptotic theory detailed by \cite{H:1996}, we expect the (limit) distribution of the test statistics considered in the following to depend on the unknown parameter $\lambda$. Consequently, we use for the implementation of the tests a similar parametric bootstrap procedure as the one suggested by \cite{GH:2000}. For a given sample $X_1, \ldots, X_n$ and a statistic $T_n$, simulate an approximate critical value $c_{n, B}^*$ for a level $\alpha$ test procedure as follows:
\begin{enumerate}
	\item[1)] Calculate $\widehat{\lambda}_n(X_1, \ldots, X_n)$ and generate $B$ bootstrap samples of size $n$ with distribution $\mathrm{Po}(\widehat{\lambda}_n)$, i.e., generate i.i.d. $\mathrm{Po}(\widehat{\lambda}_n)$ random variables $X_{j, 1}^*, \ldots, X_{j, n}^*$, $j = 1, \ldots, B$.
	
	\item[2)] Compute $T_{j, n}^* = T_n(X_{j, 1}^*, \ldots, X_{j, n}^*)$ for $j = 1, \ldots, B$.
	
	\item[3)] Derive the order statistics $T_{1:B}^* \leq \ldots \leq T_{B:B}^*$ of $T_{1, n}^*, \ldots, T_{B, n}^*$ and calculate
	\begin{equation*}
		c_{n, B}^* = T_{k:B}^* + (1 - \alpha) \cdot \big( T_{(k + 1):B}^* - T_{k:B}^* \big),
	\end{equation*}
	where $k = \lfloor (1 - \alpha) \cdot B \rfloor$ and $\lfloor \cdot \rfloor$ denotes the floor function.
	
	\item[4)] Reject the hypothesis $\mathbf{H_0}$ if $T_n(X_1, \ldots, X_n) > c_{n, B}^*$.
\end{enumerate}
This parametric bootstrap procedure was used for all of the following procedures to generate the critical points. We consider the test of \cite{BH:1992} based on the statistic
\begin{equation*}
	\mbox{BH}
	= \frac{1}{n} \sum_{i, j = 1}^n \bigg( \frac{\widehat{\lambda}_n^2}{X_i + X_j + 1} + \frac{X_i \, X_j}{X_i + X_j - 1} \bigg) - \widehat{\lambda}_n \Bigg( n - \frac{1}{n} \Bigg( \sum_{j = 1}^n \mathds{1}\{ X_j = 0 \} \Bigg)^2 \Bigg).
\end{equation*}
The mean distance test by \cite{SR:2004} is based on
\begin{equation*}
	\mbox{SR}
	= n \sum_{j = 0}^\infty \Big( \widehat{M}_n(j) - \mathrm{P}(j; \widehat{\lambda}_n) \Big)^2 \mathrm{p}(j; \widehat{\lambda}_n),
\end{equation*}
where $\widehat{M}_n(j)$ is an estimator of the CDF based on the mean distance and $\mathrm{P}(j; \lambda)$ (resp. $\mathrm{p}(j; \lambda)$) denote the distribution function (resp. pmf) of $\mathrm{Po}(\lambda)$, respectively. Note that $\mbox{SR}$ is implemented in the \texttt{R}-package \texttt{energy}, see \cite{SR:2019}. The test of \cite{ROP:1991} is based on
\begin{equation*}
	\mbox{RU}
	= \frac{1}{n} \sum_{i, j = 1}^n \frac{1}{X_i + X_j + 1} + \frac{n \cdot (1 - e^{- 2 \widehat{\lambda}_n})}{2 \, \widehat{\lambda}_n} - 2 \sum_{i = 1}^n \Bigg( \frac{(-1)^{X_i} \cdot X_i! \cdot (1 - e^{- \widehat{\lambda}_n})}{\widehat{\lambda}_n^{X_i + 1}} + \sum_{j = 1}^{X_i} \frac{(-1)^{j + 1} \cdot X_i!}{(X_i - j + 1)! \cdot \widehat{\lambda}_n^{j}} \Bigg).
\end{equation*}
Note that in the original paper of \cite{ROP:1991} and in a slight handwritten correction thereof available on the internet, as well as in the work of \cite{GH:2000}, the explicit formula of the $\mbox{RU}$-statistic contains errors. We have corrected and numerically checked the formula given above against the integral representation used to introduce the test. The integrated distribution function based tests of \cite{K:1999} are defined via
\begin{equation*}
	\mbox{K}_1
	= \sqrt{n} \Bigg( \sum_{j = 0}^M \Big| \widehat{F}_n(j) - \mathrm{P}(j; \widehat{\lambda}_n) \Big| + \widehat{\lambda}_n - \sum_{j = 0}^M \Big( 1 - \mathrm{P}(j; \widehat{\lambda}_n) \Big) \Bigg)
\end{equation*}
and
\begin{equation*}
	\mbox{K}_2
	= \sqrt{n} \sup_{1 \, \leq \, k \, \leq \, M} \Bigg| \sum_{j = 0}^{k - 1} \Big( \widehat{F}_n(j) - \mathrm{P}(j; \widehat{\lambda}_n) \Big) \Bigg|,
\end{equation*}
where $M = \max\{ X_1, \ldots, X_n \}$ and $\widehat{F}_n$ is the empirical distribution function of $X_1, \ldots, X_n$. For representations of the Kolmogorov-Smirnov statistic and the modified Cram\'{e}r-von Mises statistic, we follow the representation given by \cite{GH:2000}, namely
\begin{align*}
	\mbox{KS}
	= \sqrt{n} \sup_{0 \, \leq \, k \, \leq \, M} \Big| \widehat{F}_n(x) - \mathrm{P}(k; \widehat{\lambda}_n) \Big|
\end{align*}
and
\begin{align*}
	\mbox{CM}
	= n \sum_{j = 0}^M \Big( \widehat{F}_n(j) - \mathrm{P}(j; \widehat{\lambda}_n) \Big)^2 \cdot \frac{1}{n} \sum_{k = 1}^n \mathds{1}\{ X_k = j \}.
\end{align*}
The simulation study consists of the following 45 representatives of families of distributions. In order to show that all the considered testing procedures maintain the nominal level $\alpha$ of 5\%, we consider the Po$(\lambda)$ distribution with $\lambda\in\{1, 5, 10, 30\}$. As examples for alternative distributions, we consider the discrete uniform distribution $\mathcal{U}\{ 0, 1, \ldots, m \}$ with $m \in \{ 1, 2, 3, 5, 6 \}$, several different instances of the binomial distribution Bin$(m, q)$, several Poisson mixtures of the form PP$(q; \vartheta_1, \vartheta_2) = q \cdot \mbox{Po}(\vartheta_1) + (1 - q) \cdot \mbox{Po}(\vartheta_2)$, a $0.9/0.1$ mixture of Po$(3)$ and point mass in $0$ denoted by Po$(3)\delta_0$, discrete Weibull distributions W$(\vartheta_1, \vartheta_2)$, zero-modified Poisson distributions zmPo$(\lambda,  q)$, the zero-truncated Poisson distributions ztPo$(\lambda)$ with $\lambda \in \{2, 3, 5\}$, and the absolute discrete normal distribution $|\mbox{N}(\mu, 1)|$ with $\mu \in \{0, 2, 3\}$. Note that most distributions were generated by the packages \texttt{extraDistr}, see \cite{W:2019}, and \texttt{actuar}, see \cite{DGP:2008}, and that a significant part of these distributions can also be found in the simulation study presented by \cite{GH:2000}. Furthermore we indicate that the chosen design of simulation parameters coincides with the study by \cite{GH:2000} which facilitates the comparison to other tests of poissonity not considered here.

Every entry in Table \ref{tab:simresults} is based on 100000 repetitions and 500 bootstrap samples of size $50$. All of the considered procedures maintain the significance level $\alpha = 5\%$ under the hypothesis, which supports the statement that the parametric bootstrap procedure is well calibrated. Overall the best performing tests are K$_2$, BH and SR. The new test based on $T_n^{Po}$ is competitive to the stated procedures, although it never outperforms them all at once for the considered alternatives.
\begin{table*}[h!]
	\small
	\onehalfspacing
	\centering
	\begin{tabular}{lrrrrrrrrr}
		\toprule
		{Distr.\,\,/\,\,Test} & & \multicolumn{1}{c}{$T_n^{Po}$} & \multicolumn{1}{c}{BH} & \multicolumn{1}{c}{SR} & \multicolumn{1}{c}{RU} & \multicolumn{1}{c}{K$_1$} & \multicolumn{1}{c}{K$_2$} & \multicolumn{1}{c}{KS} & \multicolumn{1}{c}{CM} \\
		
		\cmidrule[0.4pt](r{0.125em}){1-1}
		\cmidrule[0.4pt](lr{0.125em}){3-3}
		\cmidrule[0.4pt](lr{0.125em}){4-4}
		\cmidrule[0.4pt](lr{0.125em}){5-5}
		\cmidrule[0.4pt](l{0.25em}){6-6}
		\cmidrule[0.4pt](l{0.25em}){7-7}
		\cmidrule[0.4pt](l{0.25em}){8-8}
		\cmidrule[0.4pt](l{0.25em}){9-9}
		\cmidrule[0.4pt](l{0.25em}){10-10}
		
		Po$(1)$ &  &  5 &   5 &   5 &   5 &   5 &   5 &   5 &   5 \\
		\myrowcolour
		Po$(5)$ &  &  5 &   5 &   5 &   5 &   5 &   5 &   5 &   5 \\
		Po$(10)$ &  &  5 &   5 &   5 &   5 &   5 &   5 &   5 &   5 \\
		\myrowcolour
		Po$(30)$ &  &  5 &   5 &   5 &   5 &   5 &   5 &   5 &   5 \\[1mm]
		$\mathcal{U}\{0,1\}$&  & 99 &  99 &  99 &  99 &  99 &  99 &  99 &  99 \\
		\myrowcolour
		$\mathcal{U}\{0,1,2\}$ &  &  39 &   9 &  22 &  15 &  64 &  68 &  50 &  58 \\
		$\mathcal{U}\{0,1,2,3\}$ &  &  46 &  33 &  20 &  27 &  61 &  16 &  45 &  51 \\
		\myrowcolour
		$\mathcal{U}\{0,1,2,3,4,5\}$ &  &  69 &  65 &  58 &  62 &  75 &  39 &  60 &  63 \\
		$\mathcal{U}\{0,1,2,3,4,5,6\}$ &  &  85 &  85 &  83 &  85 &  86 &  66 &  72 &  76 \\[1mm]
		\myrowcolour
		Bin$(2, 0.5)$ &  &  81 &  81 &  89 &  87 &  86 &  90 &  83 &  81 \\
		Bin$(4, 0.25)$ &  &  18 &  22 &  23 &  24 &  18 &  22 &  21 &  15 \\
		\myrowcolour
		Bin$(10, 0.1)$ &  &  7 &   7 &   7 &   7 &   6 &   7 &   7 &   6 \\
		Bin$(10, 0.5)$ &  &   57 &  52 &  49 &  52 &  82 &  88 &  60 &  68  \\
		\myrowcolour
		Bin$(1, 0.5)$ &  &  73 &  77 &  82 &  81 &  80 &  82 &  76 &  77 \\
		Bin$(2, 2/3)$ &  &  34 &  38 &  44 &  44 &  43 &  45 &  37 &  39 \\
		\myrowcolour
		Bin$(3, 0.75)$ &  &   19 &  22 &  26 &  26 &  26 &  27 &  21 &  23 \\
		Bin$(9, 0.9)$ &  &  6 &   7 &   8 &   8 &   8 &   8 &   7 &   8 \\
		\myrowcolour
		Bin$(5, 0.5)$ &  &  82 &  85 &  80 &  84 &  88 &  89 &  67 &  71 \\
		Bin$(10, 2/3)$ &  &  41 &  45 &  41 &  44 &  48 &  50 &  28 &  31 \\
		\myrowcolour
		Bin$(15, 0.75)$ &  &  23 &  27 &  26 &  26 &  28 &  30 &  16 &  18 \\
		Bin$(45, 0.9)$ &  &  8 &   9 &  10 &   9 &   9 &   8 &   6 &   7 \\[1mm]
		\myrowcolour
		PP$(0.5; 2,5)$ &  &  64 &  64 &  69 &  65 &  72 &  74 &  53 &  57 \\
		PP$(0.5; 3,5)$ &  &  17 &  19 &  20 &  19 &  20 &  21 &  12 &  14 \\
		\myrowcolour
		PP$(0.25; 1,5)$ &  &  93 &  95 &  96 &  95 &  87 &  88 &  75 &  73 \\
		PP$(0.05; 1,5)$ &  &  23 &  33 &  32 &  32 &  13 &  12 &   8 &   7 \\
		\myrowcolour
		PP$(0.01; 1,5)$ &  &  7 &   9 &   9 &   9 &   6 &   5 &   5 &   5 \\[1mm]
		Po$(3)\delta_0$ &  &  54 &  62 &  54 &  59 &  32 &  31 &  32 &  26 \\[1mm]
		\myrowcolour
		W$(0.5, 1)$ &  &  73 &  77 &  82 &  81 &  80 &  82 &  76 &  77 \\
		W$(0.25, 1)$ &  &  22 &  24 &  27 &  28 &  26 &  26 &  26 &  25 \\
		\myrowcolour
		W$(0.5, 2)$ &  &  49 &  52 &  52 &  51 &  48 &  52 &  51 &  52 \\
		W$(0.25, 2)$ &  &   8 &   8 &   7 &   6 &   6 &   8 &   7 &  10 \\
		\myrowcolour
		W$(0.75, 2)$ &  &  28 &  32 &  35 &  35 &  26 &  32 &  30 &  21 \\
		W$(0.1, 1)$ &  &  10 &  10 &  10 &   8 &  10 &  10 &  10 &  10 \\
		\myrowcolour
		W$(0.9, 3)$ &  &  97 &  97 &  99 &  99 &  98 &  99 &  97 &  93 \\[1mm]
		zmPo$(1, 0.1)$ &  &   91 &  93 &  90 &  92 &  81 &  84 &  90 &  64 \\
		\myrowcolour
		zmPo$(1, 0.5)$&  & 17 &  18 &  19 &  19 &  19 &  18 &  18 &  19 \\
		zmPo$(1, 0.8)$ &  &  71 &  72 &  74 &  74 &  74 &  74 &  74 &  74 \\
		\myrowcolour
		zmPo$(2, 0.1)$ &  &  8 &   9 &   8 &   8 &   6 &   6 &   6 &   6  \\
		zmPo$(3, 0.1)$ &  &  24 &  30 &  24 &  27 &  13 &  12 &  11 &   9 \\[1mm]
		\myrowcolour
		ztPo$(2)$ &  &  93 &  99 &  83 &  95 &  38 &  39 &  56 &  19 \\
		ztPo$(3)$ &  &  12 &  18 &  18 &  18 &   9 &  10 &   7 &   9 \\
		\myrowcolour
		ztPo$(5)$ &  &   4 &   1 &   1 &   1 &   5 &   5 &   5 &   5 \\[1mm]
		$|\mbox{N}(0, 1)|$ &  &   45 &  48 &  48 &  50 &  42 &  46 &  47 &  41 \\
		\myrowcolour
		$|\mbox{N}(2, 1)|$ &  &  44 &  46 &  59 &  53 &  54 &  61 &  42 &  37 \\
		$|\mbox{N}(3, 1)|$ &  &  88 &  78 &  94 &  86 &  96 &  98 &  85 &  90 \\
		
		\bottomrule
	\end{tabular}
	\caption{Empirical rejection rates of the tests of poissonity (sample size $n=50$, significance level $\alpha=0.05$).}\label{tab:simresults}
\end{table*}

\newpage

\section{Parameter estimation in the family of negative binomial distributions}
\label{SEC negative binomial goodness-of-fit}
\begin{figure}
\centering
\includegraphics[scale = 0.65]{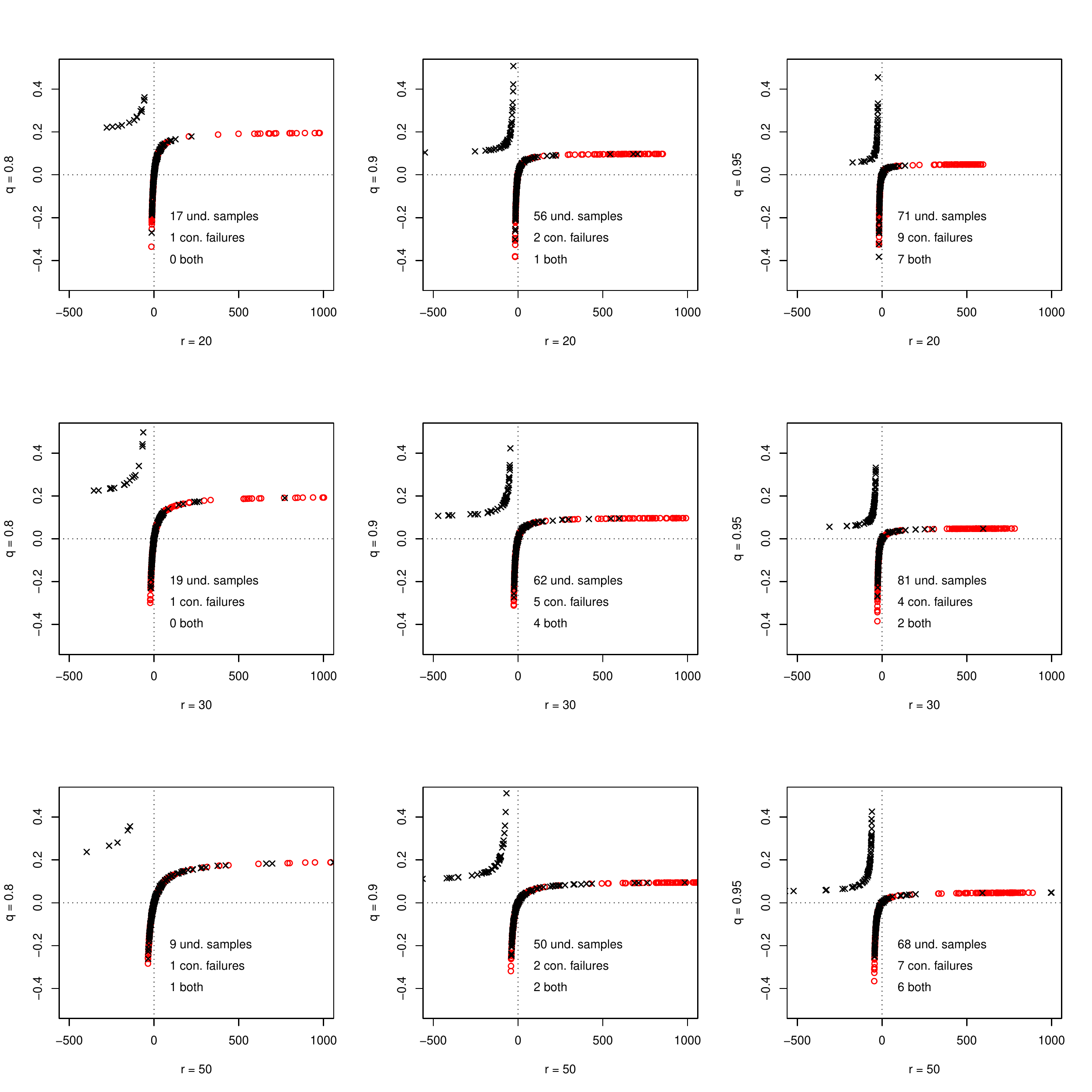}
\caption{Biases of the minimum distance (red) and method of moments (black) estimation procedures simulated for different parameters, $(r, q)\in\{20,30,50\}\times\{0.8,0.9,0.95\}$, of the negative binomial distribution with sample size $n = 100$ and $200$ repetitions of the simulations. A red circle represents the value of $(\widehat{r}_n-r_0,\widehat{q}_n-q_0)$, while a black cross stands for $(\widetilde{r}_n-r_0,\widetilde{q}_n-q_0)$. The assumed true value $(r_0,q_0)$ in every subfigure is highlighted by the intersection of the dotted lines. The number of cases of underdispersed samples, the number of convergence failures of the optimization routine, and the number of cases where both occurred are stated in the plot.} \label{fig:bias2}
\end{figure}

The characterizations we employ contain information about the underlying probability law and lead to empirical discrepancy measures being close to zero if the distribution generating the data is the one stated in the characterization. These measures can be used for estimation of the parameters of the considered parametric family of distributions. To illustrate this point, we propose a minimum distance estimation procedure for the family of negative binomial distributions. Our objective is to estimate the unknown parameters $q \in (0, 1)$ and $r > 0$ of a negative binomial distribution based on an i.i.d. $\N_0$-valued sample $X_1, \dots, X_n$. Estimation in this particular family is not trivial, since \cite{AEE:1992} have shown the conjecture of Anscombe dating back to $1950$, that the maximum likelihood equations have a unique solution if, and only if, $\overline{X}_n < S_n^2$ with $\overline{X}_n = n^{-1} \sum_{j = 1}^n X_j$, the sample mean, and $S_n^2 = n^{-1} \sum_{j = 1}^n ( X_j - \overline{X}_n )^2$, the sample variance. However, as \cite{JKK:1993} state in their Section 8.3, so called "[...] underdispersed samples [...] will occasionally be encountered, even when a negative binomial model is appropriate." The moment estimators defined by $\widetilde{q}_n = \overline{X}_n / S_n^2$ and
\begin{align*}
	\widetilde{r}_n
	= \frac{\big( \overline{X}_n \big)^2}{(1 - \widehat{q}_n) \, S_n^2}
	= \frac{\big( \overline{X}_n \big)^2}{S_n^2 - \overline{X}_n},
\end{align*}
see display (5.49) and (5.50) of \cite{JKK:1993}, perform comparably bad as the maximum likelihood estimators, since, in underdispersed samples, they lead to negative values of $\widetilde{r}_n$ or values of $\widetilde{q}_n$ that are greater than one, see the following simulation study and Figure \ref{fig:bias2}.

The heuristic for our new method is similar to that of the previous section, again based on Theorem \ref{THM chara pmf forward difference} (see also Example \ref{EXA negative binomial}). Thus, we define
\begin{align*}
	\widehat{e}_n(k;r,q) = \frac{1}{n} \sum_{j = 1}^{n} \bigg( 1 - \frac{r + X_j}{X_j + 1} \, ( 1 - q ) \bigg) \mathds{1}\{ X_j \geq k \}, \quad k \in \N_0,
\end{align*}
and let $\widehat{\rho}_n$ be as in the previous section. Similar to the test for the Poisson distribution, we consider the empirical discrepancy measure
\begin{align*}
	S_n^{NB}(r,q)
	&= \sum_{k = 0}^\infty \big( \widehat{e}_n(k;r,q) - \widehat{\rho}_n(k) \big)^2 \\
	&= \frac{1}{n^2} \sum_{j, \, \ell = 1}^n \bigg[ \bigg( 1 - \frac{r + X_j}{X_j + 1} \, ( 1 - q ) \bigg) \Big( q \big( r + X_\ell \big) - r - 1 \Big) \, \mathds{1}\{ X_j \geq X_\ell \} \\
	&\qquad\qquad\qquad + \Big( q \big( r + X_j \big) - r + 1 \Big) \bigg( 1 - \frac{r + X_\ell}{X_\ell + 1} \, \big( 1 - q \big) \bigg) \mathds{1}\{ X_j < X_\ell \} + \mathds{1}\{ X_j = X_\ell \} \bigg] ,
\end{align*}
where the proposed estimators for $(r,q)$ are defined by
\begin{equation}\label{eq:MDE}
	(\widehat{r}_n, \widehat{q}_n)
	= \mbox{argmin}_{(r, q)} S_n^{NB}(r, q).
\end{equation}
In this particular example we expect that it is possible to minimize the quadratic equation in $r$ and $q$ explicitly to obtain formulae for the estimators. However, in the following section we cannot hope for an explicit solution to the optimization problem and for reasons of consistency of the presentation, we use a numerical routine to find the values of the estimators in both cases. Note that similar estimators for parametric families of continuous distributions are investigated by \cite{BEK:2019}.

\begin{figure}[b!]
	\centering
	\includegraphics[scale = 0.65]{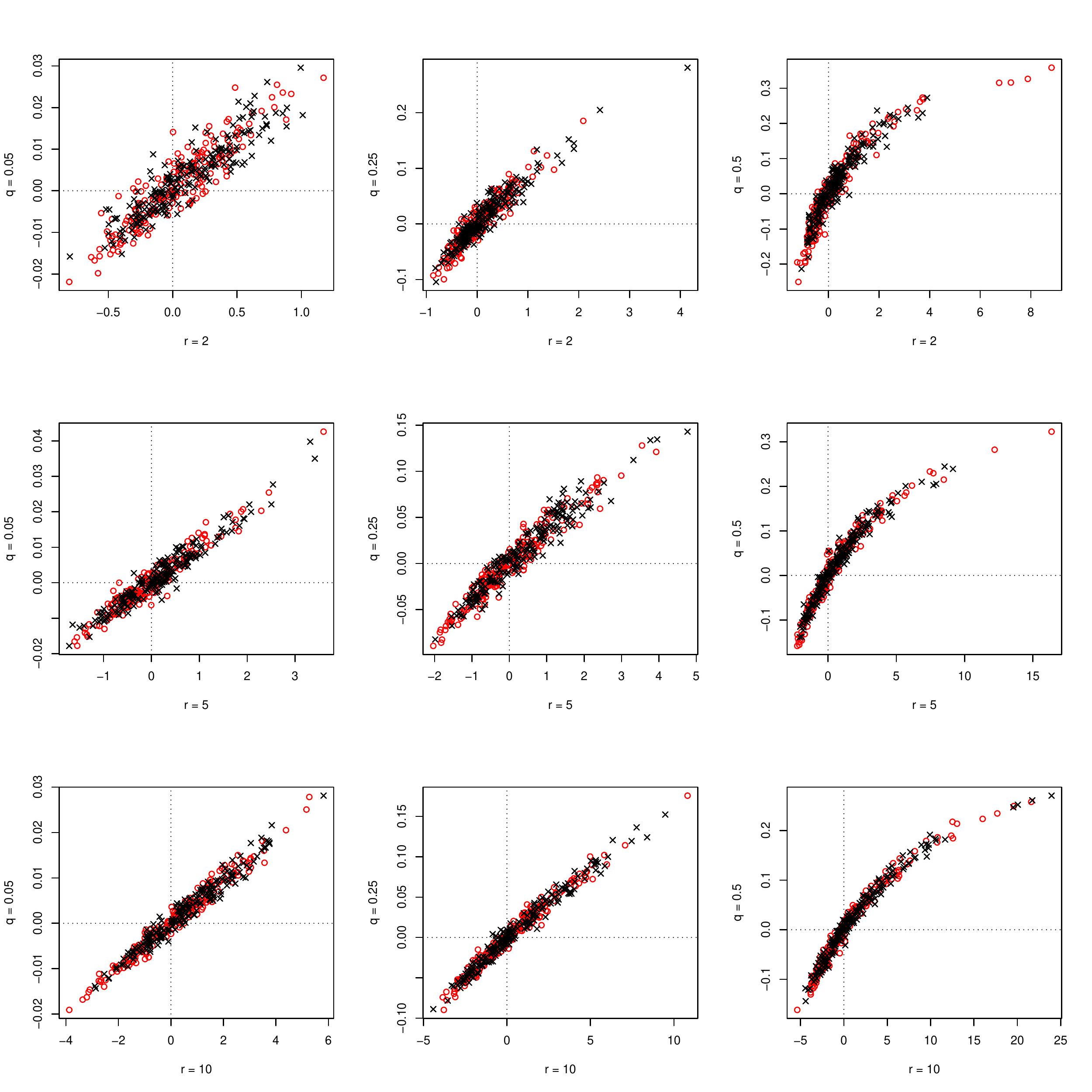}
	\caption{Biases of the minimum distance (red) and method of moments (black) estimation procedures simulated for different parameters, $(r, q)\in\{2,5,10\}\times\{0.05,0.25,0.5\}$, of the negative binomial distribution with sample size $n = 100$ and $200$ repetitions of the simulations. A red circle represents the value of $(\widehat{r}_n-r_0,\widehat{q}_n-q_0)$, while a black cross stands for $(\widetilde{r}_n-r_0,\widetilde{q}_n-q_0)$. The assumed true value $(r_0,q_0)$ in every subfigure is highlighted by the intersection of the dotted lines.} \label{fig:bias1}
\end{figure}

For a comparison of the two presented methods we conduct a simulation study in \texttt{R} and use the \texttt{optim} routine to find the minimal values in (\ref{eq:MDE}). The option \texttt{method} was fixed to \texttt{L-BFGS-B}, thus choosing an implementation of the routine suggested by \cite{BLNZ:1995}, and the maximum number of iterations to \texttt{maxit=1000}. As starting values for the optimization routine we choose independent uniformly distributed random numbers $r_{s} \sim \mathcal{U}(1, 3)$ and $q_s \sim \mathcal{U}(0.1, 0.9)$. For different $(r_0, q_0) \in (0, \infty) \times (0, 1)$ we simulate $200$ i.i.d. samples of size $n = 100$ from a negative binomial distribution with parameters $(r_0, q_0)$ and calculate the minimum distance estimators $(\widehat{r}_n, \widehat{q}_n)$ as well as the moment estimators $(\widetilde{r}_n, \widetilde{q}_n)$. Then the bias of the estimation is derived by subtracting the underlying 'true' parameters $(r_0, q_0)$. In Figures \ref{fig:bias1} and \ref{fig:bias2} the results of the different simulations are plotted, with estimation results of the moment estimators plotted as black crosses and the results of the minimum distance estimators as red circles. It is visible in Figure \ref{fig:bias1} that for small values of $q_0$, both procedures perform comparably, although the values of the moments estimators seem to scatter a little more than those of the minimum distance estimators. A completely different picture is seen in Figure \ref{fig:bias2}, where values of $q_0$ in the neighborhood of $1$ and greater values of $r_0$ are assumed. The moment estimators $(\widetilde{r}_n, \widetilde{q}_n)$ regularly produce values which are clearly outside of the defined parameter space $(0, \infty) \times (0, 1)$ as opposed to the minimum distance estimators $(\widehat{r}_n, \widehat{q}_n)$ which do not show this behavior due to the optimization constraints. Nevertheless, some convergence failures in the optimization routine did occur and they are not exclusively related to the underdispersed samples and only happen for somewhat extreme parameter configurations. We chose to visually assess the quality of estimation, since empirical versions of the bias and mean squared error are very sensitive to big discrepancies, and hence did not provide valuable information on the quality of the estimation procedures. It would be of interest to find theoretical statements for the estimators $(\widehat{r}_n, \widehat{q}_n)$ such as consistency results or a central limit theorem type asymptotic distribution.

\section{Parameter estimation in discrete exponential-polynomial models}
\label{SEC discrete exponential-polynomial parameter estimation}

In this final section we present an application to a non-normalized model, namely parameter estimation in the discrete exponential-polynomial models introduced in Example \ref{EXA exponential polynomial}. We follow \cite{BEK:2019} who apply the continuous version of our estimation method to continuous exponential-polynomial models. In their work, they compare the method with two other methods for parameter estimation in non-normalized continuous models. More specifically, they implemented the score matching approach of \cite{H:2007} as well as noise-contrastive estimators from \cite{GH:2012}. As another contribution that focuses on the continuous exponential-polynomial distribution and the corresponding parameter estimation problem, let us mention \cite{HT:2016}. In our search through the literature we have found only few methods for the parameter estimation in the discrete version of the model. As such, contrastive divergence methods based on the initial proposal by \cite{H:2002} can be applied in principle though it does not avoid dealing with the normalization constant $C(\vartheta)$ and \cite{Ly:2009} proposes a discrete version of the score matching approach but does not give details on its implementation.
More recently \cite{TK:2017} proposed a method (that avoids any calculation or approximation of the normalization constant) based on suitable homogeneous divergences which are empirically localized.
We use this latter method as a comparison to our approach.

Assume that $X_1, \dots, X_n$ is an i.i.d. $\N$-valued sample from the exponential-polynomial model $p_{\vartheta^{(0)}}$ in Example \ref{EXA exponential polynomial} with some unknown parameter $\vartheta^{(0)} \in \R^{d - 1} \times (- \infty, 0)$ (with $d \in \N$, $d \geq 2$, fixed and known). We seek to estimate $\vartheta^{(0)}$ based on $X_1, \dots, X_n$. Very similar to the previous section, we consider $\widehat{\rho}_n$ as before, and put
\begin{align*}
	\widehat{e}_n(k ; \vartheta)
	= \frac{1}{n} \sum_{j = 1}^n \bigg( 1 - \exp\Big( \vartheta_1 + \vartheta_2 \big( (X_j + 1)^2 - X_j^2 \big) + \dotso + \vartheta_d \big( (X_j + 1)^d - X_j^d \big) \Big) \bigg) \mathds{1}\{ X_j \geq k \}, \quad k \in \N.
\end{align*}
We define the empirical discrepancy measure
\begin{align*}
	S_n^{PE}(\vartheta)
	= \sum_{k = 0}^\infty \big( \widehat{e}_n(k; \vartheta) - \widehat{\rho}_n(k) \big)^2 .
\end{align*}
In line with Theorem \ref{THM chara pmf forward difference}, or more precisely, Example \ref{EXA exponential polynomial}, we propose as an estimator
\begin{equation}\label{eq:exppoly1}
	\widehat{\vartheta}_n
	= \mbox{argmin}_{\vartheta} S_n^{PE}(\vartheta) .
\end{equation}
To see if this approach leads to sensible estimators, we conduct simulations in a two-parameter special case of the model. Following the continuous-case simulation setting of \cite{BEK:2019}, we consider $d = 3$ but fix $\vartheta_2 = 0$, thus effectively estimating the parameters of the parametric family given through
\begin{equation}\label{eq:exppoly2}
	p_{(\vartheta_1, \vartheta_3)}(k)
	= C(\vartheta_1, \vartheta_3)^{-1} \exp\big( \vartheta_1 k + \vartheta_3 k^3 \big), \quad k \in \N, \quad \vartheta_1 \in \R, ~ \vartheta_3 < 0 ,
\end{equation}
which, though simpler than the general case, is still a non-normalized model and thus inaccessible to explicit maximum likelihood estimation. The discrepancy measure $S_n^{PE}(\vartheta)$ can be calculated as
\begin{align*}
	S_n^{PE}(\vartheta)
	= \frac{1}{n^2} \sum_{j, \ell = 1}^n \Big[& \big( E_j(\vartheta_1, \vartheta_3) - 1 \big) \big( E_\ell(\vartheta_1, \vartheta_3) \cdot X_\ell - X_\ell + 2 \big) \mathds{1}\{ X_j \geq X_\ell \} + \mathds{1}\{ X_j = X_\ell \} \\
	&+ \big( E_j(\vartheta_1, \vartheta_3) - 1 \big) \big( E_\ell(\vartheta_1, \vartheta_3) - 1 \big) \cdot X_j \cdot \mathds{1}\{ X_j < X_\ell \} \Big] ,
\end{align*}
where
\begin{align*}
	E_i (\vartheta_1, \vartheta_3)
	= \exp\big( \vartheta_1 + \vartheta_3 + 3 \vartheta_3 X_i + 3 \vartheta_3 X_i^2 \big), \quad i = 1, \dots, n .
\end{align*}

For a comparison, we consider the estimator proposed by \cite{TK:2017}. For positive constants $\alpha,\alpha',\gamma>0$, with $\alpha>\alpha'$, and $\bar{\alpha} = (\alpha + \gamma\alpha') \, / \, (1+\gamma)$, their estimator is given as
\begin{align}\label{eq:takenouchi}
\widetilde{\vartheta}_n = \mbox{argmin}_{\vartheta} \Bigg\{&\frac{1}{1+\gamma}\log\Bigg(\sum_{k\in\mathcal{Z}} \bigg(\frac{n_k}{n}\bigg)^\alpha q_\vartheta(k)^{1-\alpha}\Bigg)
+\frac{\gamma}{1+\gamma}\log\Bigg(\sum_{k\in\mathcal{Z}} \bigg(\frac{n_k}{n}\bigg)^{\alpha'} q_\vartheta(k)^{1-\alpha'}\Bigg)\nonumber\\
&-\log\Bigg(\sum_{k\in\mathcal{Z}} \bigg(\frac{n_k}{n}\bigg)^{\bar{\alpha}} q_\vartheta(k)^{1-\bar{\alpha}}\Bigg)\Bigg\},
\end{align}
where $\mathcal{Z}$ is the set of all values that appear in the sample $X_1,\ldots,X_n$, the variable $n_k$ denotes how often the value $k$ is found in the sample, and
$$
q_\vartheta(k) = \exp\big(\vartheta_1 k +\ldots+\vartheta_d k^d \big), \quad k\in\N.
$$
Since \cite{TK:2017} do not propose a specific way of choosing the constants $\alpha$, $\alpha'$ and $\gamma$, we use the values that appear most frequently in their simulation study and therefore set $\alpha=1.1$, $\alpha'=0.1$ and $\gamma= 1/9$.

\begin{figure}[h]
\centering
\includegraphics[scale = 0.65]{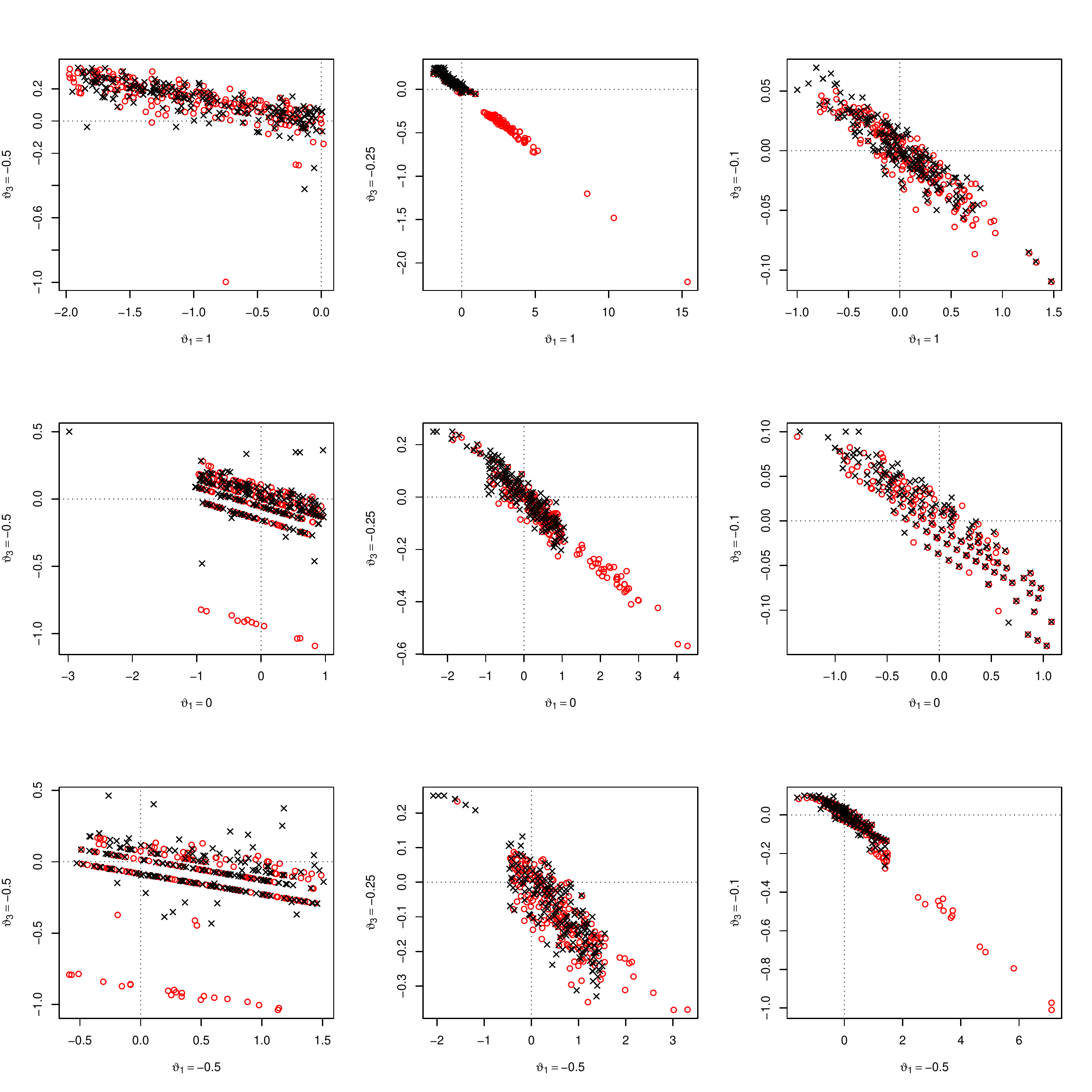}
\caption{Simulated biases of the estimators $\widehat{\vartheta}_n$ (red) and $\widetilde{\vartheta}_n$ (black) in the discrete exponential-polynomial model for different parameters $\big(\vartheta_1^{(0)},\vartheta_3^{(0)}\big)$ ($n=100$; 200 repetitions).
A red circle represents the value of ${\big(\widehat{\vartheta}_{n,1} - \vartheta_1^{(0)}, \widehat{\vartheta}_{n,3}- \vartheta_3^{(0)}\big)}$, while a black cross stands for ${\big(\widetilde{\vartheta}_{n,1} - \vartheta_1^{(0)}, \widetilde{\vartheta}_{n,3}- \vartheta_3^{(0)}\big)}$. The assumed true value  in every subfigure is highlighted by the intersection of the dotted lines.} \label{fig:exppoly}
\end{figure}

As in the previous section, we use the software \texttt{R} for the simulation and the \texttt{optim} routine to find the minimal values in \eqref{eq:exppoly1} and \eqref{eq:takenouchi}. Again, the option \texttt{method} is fixed to \texttt{L-BFGS-B} and the maximum number of iterations to \texttt{maxit=1000}.
As starting values for the optimization we choose independent uniformly distributed random numbers $\vartheta_{1}^{(s)} \sim \mathcal{U}(-1,1)$ and $\vartheta_{3}^{(s)} \sim \mathcal{U}(-1,0)$. For different $\big(\vartheta_{1}^{(0)}, \vartheta_{3}^{(0)}\big) \in \R \times (-\infty, 0)$, we simulate $200$ i.i.d. samples of size $n = 100$ from the discrete exponential-polynomial model in \eqref{eq:exppoly2} with parameters $(\vartheta_{1}^{(0)}, \vartheta_{3}^{(0)})$ and calculate the estimators $\big(\widehat{\vartheta}_{n,1}, \widehat{\vartheta}_{n,3}\big)$ and $\big(\widetilde{\vartheta}_{n,1}, \widetilde{\vartheta}_{n,3}\big)$ presented in \eqref{eq:exppoly1} and \eqref{eq:takenouchi} respectively. The biases of the estimators are given by subtracting the underlying 'true' parameters $\big(\vartheta_{1}^{(0)}, \vartheta_{3}^{(0)}\big)$. The simulation of a discrete exponential-polynomial model is rather simple as $\vartheta_{3}^{(0)} < 0$ ensures that the probability $p_{(\vartheta_{1}^{(0)}, \vartheta_{3}^{(0)})}(k)$ is rapidly decreasing as $k$ grows. From a practical point of view and minding the usual calculation accuracy, we only need to deal with a discrete distribution with finite support. In Figure \ref{fig:exppoly} the results of the simulations are presented and it is visible that both procedures perform comparable and overall well. The newly proposed estimators tend to scatter more which favors the competing estimators. However, our new estimators require no (data-dependent or quick fix) choice of parameters \citep[like $\alpha$, $\alpha'$ and $\gamma$ for the estimators of][]{TK:2017}. Introducing additional parameters, for instance through suitable weight functions, is also conceivable for our method. It would certainly allow for some choice which improves the overall performance, but it also leads to a less intuitive implementation as these parameters need to be chosen in practice. Note that, as in the previous simulation, some convergence failures in the optimization routine occurred (less than ten percent per parameter configuration).

\vspace{5mm}
\appendix
\Large
\textbf{Appendix}
\normalsize

\section{Proof of Theorem \ref{THM density approach}}
\label{APP SEC proof of density approach}

The following proof is, up to technical details involving the class of test functions, due to \cite{LS:2013} and given here for the reader's convenience.

Assume that $X|p \sim p$. Then, for $f \in \mathcal{F}_p$,
\begin{align*}
\E \left[ \Delta^+ f(X) + \frac{\Delta^+ p(X)}{p(X)} \, f(X + 1) ~ \bigg| ~ X \in \mathrm{spt}(p) \right]
&= \sum_{k = L}^R \Big( p(k) \, \Delta^+ f(k) + f(k + 1) \, \Delta^+ p(k) \Big) \\
&= \sum_{k = L}^R \Delta^+ \big( p(k) \, f(k) \big) 
= 0,
\end{align*}
using assumption ($a$). To prove the converse, take a discrete random variable $Z$ with mass function $p$, independent of $X$. For $m \in \Z$, define $f_m : \big\{ L, \dots, R\big\} \to \R$ via
\begin{align*}
f_m(k)
= \frac{1}{p(k)} \sum_{\ell = L}^{k - 1} \big( \mathds{1}\{ \ell \leq m \} - \PP(Z \leq m) \big) \, p(\ell)
\end{align*}
which satisfies
\begin{align*}
\Delta^+ \big( p(k) \, f_m(k) \big)
= \big( \mathds{1}\{ k \leq m \} - \PP(Z \leq m) \big) \, p(k), \quad k \in \{ L, \dots, R \}.
\end{align*}
Therefore,
\begin{align*}
\sum_{k = L}^R \Big| \Delta^+ \big( p(k) \, f_m(k) \big) \Big|
\leq 2 \sum_{k = L}^R p(k)
= 2
< \infty,
\end{align*}
as well as
\begin{align*}
\sum_{k = L}^R \Delta^+ \big( p(k) \, f_m(k) \big)
= \sum_{k = L}^R \big( \mathds{1}\{ k \leq m \} - \PP(Z \leq m) \big) \, p(k)
= \PP(Z \leq m) - \PP(Z \leq m)
= 0.
\end{align*}
Moreover, we have $f_m(k + 1) \leq 2 \cdot P(k) \, / \, p(k + 1)$ and
\begin{align*}
	\big| f_m(k + 1) \big|
	&= \frac{1}{p(k + 1)} \, \Bigg| \sum_{\ell = L}^{R} \big( \mathds{1}\{ \ell \leq m \} - \PP(Z \leq m) \big) \, p(\ell) - \sum_{\ell = k + 1}^{R} \big( \mathds{1}\{ \ell \leq m \} - \PP(Z \leq m) \big) \, p(\ell) \Bigg| \\
	&= \frac{1}{p(k + 1)} \, \Bigg| \sum_{\ell = k + 1}^{R} \big( \mathds{1}\{ \ell \leq m \} - \PP(Z \leq m) \big) \, p(\ell) \Bigg| \\
	&\leq \frac{2}{p(k + 1)} \cdot \big( 1 - P(k) \big) ,
\end{align*}
and thus get
\begin{align*}
\sup_{k \in \{ L, \dots, R \}} \left| \frac{\Delta^{+} p(k)}{p(k)} \, f_m(k + 1) \right|
\leq 2 \cdot \sup_{k \in \{ L, \dots, R - 1 \}} \left| \frac{\Delta^{+} p(k) \cdot \min\{ P(k), 1 - P(k) \}}{p(k) \, p(k + 1)} \right|
< \infty .
\end{align*}
Now, notice that
\begin{align*}
\Delta^+ f_m(k)
&= \left( \frac{1}{p(k + 1)} - \frac{1}{p(k)} \right) \sum_{\ell = L}^{k} \big( \mathds{1}\{ \ell \leq m \} - \PP(Z \leq m) \big) \, p(\ell) + \frac{1}{p(k)} \big( \mathds{1}\{ k \leq m \} - \PP(Z \leq m) \big) \, p(k) \\
&= \left( 1 - \frac{p(k + 1)}{p(k)} \right) \frac{1}{p(k + 1)} \sum_{\ell = L}^{k} \big( \mathds{1}\{ \ell \leq m \} - \PP(Z \leq m) \big) \, p(\ell) + \mathds{1}\{ k \leq m \} - \PP(Z \leq m) \\
&= - \frac{\Delta^+ p(k)}{p(k)} \, f_m(k + 1) + \mathds{1}\{ k \leq m \} - \PP(Z \leq m) ,
\end{align*}
where the calculation is valid for $k \in \{ L, \dots, R - 1 \}$, but the equality obviously also holds for $k = R$ [using our convention $f_m(R + 1) = 0$] if $R < \infty$. From this relation, we immediately get that
\begin{align*}
\sup_{k \in \{ L, \dots, R \}} \big| \Delta^+ f_m(k) \big| < \infty,
\end{align*}
so $f_m \in \mathcal{F}_p$, as well as, by the assumption in the converse implication,
\begin{align*}
0
= \E \left[ \Delta^+ f(X) + \frac{\Delta^+ p(X)}{p(X)} \, f(X + 1) ~ \bigg| ~ X \in \mathrm{spt}(p) \right]
= \E \Big[ \mathds{1}\{ X \leq m \} - \PP(Z \leq m) ~ \Big| ~ X \in \mathrm{spt}(p) \Big],
\end{align*}
which implies the claim.

\section{Proof of Theorem \ref{THM chara pmf forward difference}}
\label{APP SEC proof of pmf chara}

First assume that $X|p \sim p$. Then, for all $k \in \Z$, $k \geq L$, we have
\begin{align*}	
	\rho_{X|p}(k)
	= p(k)
	= - \sum_{\ell = k}^R \big( p(\ell + 1) - p(\ell) \big)
	= - \sum_{\ell = k}^R \frac{\Delta^+ p(\ell)}{p(\ell)} \, p(\ell)
	= \E \bigg[ - \frac{\Delta^+ p(X)}{p(X)} \, \mathds{1}\{ X \geq k \} ~ \bigg| ~ X \in \mathrm{spt}(p) \bigg].
\end{align*}
For the converse implication, assume that
\begin{align*}
	\rho_{X|p}(k) = \E \bigg[ - \frac{\Delta^+ p(X)}{p(X)} \, \mathds{1}\{ X \geq k \} ~ \bigg| ~ X \in \mathrm{spt}(p) \bigg], \quad k \in \Z, k \geq L.
\end{align*}
We obtain for $f \in \mathcal{F}_p$
\begin{align*}
	\E \Big[ \Delta^+ f(X) \, \Big| \, X \in \mathrm{spt}(p) \Big]
	= \sum_{\ell = L}^R \big( \Delta^+ f(\ell) \big) \, \rho_{X|p}(\ell)
	&= \sum_{\ell = L}^R \Delta^+ f(\ell) \cdot \E \bigg[ - \frac{\Delta^+ p(X)}{p(X)} \, \mathds{1}\{ X \geq \ell \} ~ \bigg| ~ X \in \mathrm{spt}(p) \bigg] \\
	&= \E \Bigg[ - \frac{\Delta^+ p(X)}{p(X)} \sum_{\ell = L}^X \big( f(\ell + 1) - f(\ell) \big) ~ \bigg| ~ X \in \mathrm{spt}(p) \Bigg] \\
	&= \E \bigg[ - \frac{\Delta^+ p(X)}{p(X)} \, f(X + 1) ~ \bigg| ~ X \in \mathrm{spt}(p) \bigg],
\end{align*}
where we use that $f(L) = 0$ by Remark \ref{RMK test functions}, and where Fubini's theorem is applicable as
\begin{align*}
	\sum_{\ell = L}^R \E \bigg| \Delta^+ f(\ell) \, \frac{\Delta^+ p(X)}{p(X)} \, \mathds{1}\big\{ X \geq \ell, \, X \in \mathrm{spt}(p) \big\} \bigg|
	&\leq \sup_{k \in \{ L, \dots, R \}} \big| \Delta^+ f(k) \big| \cdot \E \bigg| \frac{\Delta^+ p(X)}{p(X)} \, (X - L) \cdot \mathds{1}\big\{ X \in \mathrm{spt}(p) \big\} \bigg| \\
	&< \infty.
\end{align*}
Theorem \ref{THM density approach} implies the claim.

\section{Proof of Proposition \ref{PROP chara distribution functions}}
\label{APP SEC proof of distribution function chara}

If $X|p \sim p$, then Theorem \ref{THM chara pmf forward difference} implies that, for $k \in \Z$, $k \geq L$,
\begin{align*}
	F_{X|p}(k)
	= \sum_{\ell = L}^k \rho_{X|p}(\ell)
	&= \E \Bigg[ - \frac{\Delta^+ p(X)}{p(X)} \sum_{\ell = L}^k \mathds{1}\{ X \geq \ell \} ~ \bigg| ~ X \in \mathrm{spt}(p) \Bigg] \\
	&= \E \bigg[ - \frac{\Delta^+ p(X)}{p(X)} \big( \min\{ X, k \} - L + 1 \big) ~ \bigg| ~ X \in \mathrm{spt}(p) \bigg].
\end{align*}
To prove the converse, assume that
\begin{align*}
	F_{X|p}(k) = \E \bigg[ - \frac{\Delta^+ p(X)}{p(X)} \big( \min\{ X, k \} - L + 1 \big) ~ \bigg| ~ X \in \mathrm{spt}(p) \bigg] , \quad k \in \Z, k \geq L.
\end{align*}
Then, we have
\begin{align*}
	\rho_{X|p}(k)
	= F_{X|p}(k) - F_{X|p}(k - 1)
	&= \E \Bigg[ - \frac{\Delta^+ p(X)}{p(X)} \Bigg( \sum_{\ell = L}^k \mathds{1}\{ X \geq \ell \} - \sum_{\ell = L}^{k - 1} \mathds{1}\{ X \geq \ell \} \Bigg) ~ \bigg| ~ X \in \mathrm{spt}(p) \Bigg] \\
	&= \E \bigg[ - \frac{\Delta^+ p(X)}{p(X)} \, \mathds{1}\{ X \geq k \} ~ \bigg| ~ X \in \mathrm{spt}(p) \bigg], \quad k \in \Z, k > L,
\end{align*}
and
\begin{align*}
	\rho_{X|p}(L)
	= F_{X|p}(L)
	= \E \bigg[ - \frac{\Delta^+ p(X)}{p(X)} \, \mathds{1}\{ X \geq L \} ~ \bigg| ~ X \in \mathrm{spt}(p) \bigg],
\end{align*}
so Theorem \ref{THM chara pmf forward difference} yields $X|p \sim p$.

\section{Proof of Proposition \ref{PROP chara characteristic functions}}
\label{APP SEC proof of characteristic function chara}

First, if $X|p \sim p$, Theorem \ref{THM chara pmf forward difference} gives
\begin{align*}
	\varphi_{X|p}(t)
	= \sum_{\ell = L}^{R} e^{i t \ell} \rho_{X|p}(\ell)
	&= \E \Bigg[ - \frac{\Delta^+ p(X)}{p(X)} \sum_{\ell = L}^{X} e^{i t \ell} ~ \bigg| ~ X \in \mathrm{spt}(p) \Bigg] \\
	&= \E \bigg[ - \frac{\Delta^+ p(X)}{p(X)} \cdot \frac{e^{i t L} - e^{i t (X + 1)}}{1 - e^{i t}} ~ \bigg| ~ X \in \mathrm{spt}(p) \bigg] , \quad t \in \R,
\end{align*}
where the use of Fubini's theorem (in case $R = \infty$) is admissible by the argument from Appendix \ref{APP SEC proof of pmf chara}, and where we use the complex geometric sum in the last step. For the converse, assume that $\varphi_{X|p}$ is given through the stated formula. The inversion formula for characteristic functions applied to the atoms of $X|X \in \mathrm{spt}(p)$ \citep[see Theorem 2.3.2 of][]{C:1975} gives, for $k \in \Z$, $k \geq L$,
\begin{align*}
	\rho_{X|p}(k)
	&= \lim_{T \to \infty} \frac{1}{2 T} \int_{-T}^{T} e^{- i t k} \varphi_{X|p}(t) \, \mathrm{d}t \\
	&= \lim_{T \to \infty} \frac{1}{2 T} \int_{-T}^{T} e^{- i t k} \, \E \bigg[ - \frac{\Delta^+ p(X)}{p(X)} \cdot \frac{e^{i t L} - e^{i t (X + 1)}}{1 - e^{i t}} ~ \bigg| ~ X \in \mathrm{spt}(p) \bigg] \mathrm{d}t \\
	&= \lim_{T \to \infty} \frac{1}{2 T} \int_{-T}^{T} e^{- i t k} \, \E \Bigg[ - \frac{\Delta^+ p(X)}{p(X)} \sum_{\ell = L}^{X} e^{i t \ell} ~ \bigg| ~ X \in \mathrm{spt}(p) \Bigg] \mathrm{d}t \\
	&= \lim_{T \to \infty} \frac{1}{2 T} \, \E \Bigg[ - \frac{\Delta^+ p(X)}{p(X)} \sum_{\ell = L}^{X} \int_{-T}^{T} e^{- i t k} e^{i t \ell} \, \mathrm{d}t ~ \bigg| ~ X \in \mathrm{spt}(p) \Bigg] \\
	&= \lim_{T \to \infty} \frac{1}{2 T} \, \E \Bigg[ - \frac{\Delta^+ p(X)}{p(X)} \sum_{\ell = L}^{X} \bigg( \frac{e^{i T (\ell - k)} - e^{- i T (\ell - k)}}{i ( \ell - k)} \, \mathds{1}\{ \ell \neq k \} + 2 T \, \mathds{1}\{ \ell = k \} \bigg) ~ \bigg| ~ X \in \mathrm{spt}(p) \Bigg] \\
	&= \lim_{T \to \infty} \E \Bigg[ - \frac{\Delta^+ p(X)}{p(X)} \sum_{\ell = L}^{X} \bigg( \frac{\sin\big( T (\ell - k) \big)}{T (\ell - k)} \, \mathds{1}\{ \ell \neq k \} + \mathds{1}\{ \ell = k \} \bigg) ~ \bigg| ~ X \in \mathrm{spt}(p) \Bigg] \\
	&= \E \Bigg[ - \frac{\Delta^+ p(X)}{p(X)} \sum_{\ell = L}^{X} \mathds{1}\{ \ell = k \} ~ \bigg| ~ X \in \mathrm{spt}(p) \Bigg] \\
	&= \E \bigg[ - \frac{\Delta^+ p(X)}{p(X)} \, \mathds{1}\{ X \geq k \} ~ \bigg| ~ X \in \mathrm{spt}(p) \bigg],
\end{align*}
where the use of dominated convergence in the second to last step is easily justified (if necessary) by the integrability assumptions on $X$. Theorem \ref{THM chara pmf forward difference} yields the claim.

\section{Proof of Proposition \ref{PROP chara probability generating functions}}
\label{APP SEC proof of probability generating function chara}

The necessity part follows from Theorem \ref{THM chara pmf forward difference} with a calculation similar to Appendix \ref{APP SEC proof of characteristic function chara}. For the sufficiency part, assume that $G_{X}$ is given through the stated formula. Then, for all $s \in [0, 1)$, we have
\begin{align*}
	G_{X}(s)
	= \E \bigg[ - \frac{\Delta^+ p(X)}{p(X)} \cdot \frac{1 - s^{X + 1}}{1 - s} \bigg]
	= \sum_{\ell = 0}^\infty s^\ell \, \E\bigg[ - \frac{\Delta^+ p(X)}{p(X)} \, \mathds{1}\{ X \geq \ell \} \bigg].
\end{align*}
Since $G_X(s) = \sum_{\ell = 0}^\infty s^\ell \rho_X(\ell) \leq 1$, for $s \in [0, 1)$, $G_X$ is a convergent power series in $[0, 1)$. We can thus differentiate $G_X$ in $(0, 1)$ to obtain
\begin{align*}
	\frac{\mathrm{d}^k}{\mathrm{d} s^k} \, G_X(s)
	= \sum_{\ell = k}^\infty \ell (\ell - 1) \cdots (\ell - k + 1) \, s^{\ell - k} \rho_X(\ell)
\end{align*}
as well as
\begin{align*}
	\frac{\mathrm{d}^k}{\mathrm{d} s^k} \, G_X(s)
	= \sum_{\ell = k}^\infty \ell (\ell - 1) \cdots (\ell - k + 1) \, s^{\ell - k} \, \E\bigg[ - \frac{\Delta^+ p(X)}{p(X)} \, \mathds{1}\{ X \geq \ell \} \bigg].
\end{align*}
We conclude that
\begin{align*}
	k! \cdot \E\bigg[ - \frac{\Delta^+ p(X)}{p(X)} \, \mathds{1}\{ X \geq k \} \bigg]
	= \frac{\mathrm{d}^k}{\mathrm{d} s^k} G_X(s)\big|_{s \, = \, 0}
	= k! \cdot \rho_X(k)
\end{align*}
so Theorem \ref{THM chara pmf forward difference} implies the claim.

\vspace{5mm}

\bibliography{bib_discrete}
\bibliographystyle{apalike}

\end{document}